\documentclass[a4paper,11pt]{article}
\pdfoutput=1 

\usepackage{jcappub} 
\usepackage{amsmath}
\usepackage{eqnarray,amsmath}
\usepackage{mathtools}
\usepackage{pdfpages}
\usepackage{lipsum}
\usepackage{subfig}
\usepackage[T1]{fontenc} 
\usepackage{xcolor}

\title{\bf Revisiting coupled CDM-massive neutrino perturbations in diverse cosmological backgrounds}

\author[a,1]{Sourav Pal,\note{Corresponding author.}}
\author[b]{Rickmoy Samanta}
\author[a,c]{and Supratik Pal}

\emailAdd{soupal1729@gmail.com}
\emailAdd{rickmoysamanta@gmail.com}
\emailAdd{supratik@isical.ac.in}

\affiliation[a]{Physics and Applied Mathematics Unit, Indian Statistical Institute,\\ 203, B.T. Road, Kolkata 700108, India}
\affiliation[b]{Department of Physics, Birla Institute of Technology and Science Pilani,\\
Hyderabad 500078, India}
\affiliation[c]{Technology Innovation Hub on Data Science, Big Data Analytics and Data Curation, \\ Indian Statistical Institute, 203, B.T. Road, Kolkata 700108, India}

\abstract
{
 Massive neutrinos are well-known to cause a characteristic suppression in the growth of structures at scales below the neutrino free-streaming length. A detailed understanding of this suppression is  essential in the  era of precision cosmology we are entering into, enabling us to better constrain the total neutrino mass and possibly probe (beyond)-$\Lambda$CDM cosmological model(s). Instead of the usual N-body simulation or Boltzmann solver, in this article we consider a two-fluid framework at the linear scales, where the neutrino fluid perturbations are coupled to the CDM (+ baryon) fluid via gravity at redshifts of interest. Treating the neutrino mass fraction $f_\nu$ as a perturbative parameter, we find solutions to the system  with redshift-dependent neutrino free-streaming length in $\Lambda$CDM background via two separate approaches. The perturbative scale-dependent solution is shown to be in excellent agreement with numerical solution of the two-fluid equations valid to all orders in $f_{\nu}$, and also agrees with results from {\texttt{CLASS}} to a good accuracy. We further generalize the framework to incorporate different evolving dark energy backgrounds and found sub-percent level differences in the suppression, all of which lie within the observational uncertainty of BOSS-like surveys. We also present a brief discussion on the prospects of the current analysis in the context of upcoming missions. 
}

\begin{document}
\maketitle
\flushbottom

\section{Introduction}

The $\Lambda$CDM model, consisting of a cosmological constant ($\Lambda$) and cold dark matter (CDM), is widely accepted  as the concordance model of cosmology, not only because of the simplicity of the model but also due to its empirical success \cite{Planck:2018vyg,BOSS:2016wmc} so far as major observational data is concerned. In the baseline $\Lambda$CDM model, the universe is described by 6 parameters only, where cosmological constant sits in the background and CDM (+ baryons) perturbs as a single fluid in this background. However, apart from $\Lambda$ and CDM sectors, there are moderate to strong evidences of some other cosmic species that may take part in both background evolution and perturbations at specific eras. This in turn may reflect upon the cosmological parameters, some of which are degenerate with those representing other cosmic species. For example, as a simple extension beyond $\Lambda$CDM with massive neutrinos, some fundamental parameters like the Hubble parameter $H_0$ get affected due to possible degeneracies with the effective number of neutrino species $N_{\rm eff}$ and the total neutrino mass $\sum m_\nu$  \cite{Bernal:2016gxb}. Although the existence of massive neutrinos have been proved beyond doubt by neutrino oscillation experiments \cite{LSND:2001aii,LSND:1995lje}, it only helps in determining the mass difference between two species. We are yet to figure out the exact masses of individual neutrino species or the number of sterile neutrinos in the universe \cite{PTOLEMY:2019hkd}. The above reasons prompt us to think beyond just CDM perturbations and to find out the effects of neutrino perturbations on CDM sector and hence on cosmological observations. This may in turn help us have more precise information on forthcoming observations. On top of that, since for the dark energy (DE) sector, $\Lambda$ is the simplest choice that has its own limitations \cite{Perivolaropoulos:2021jda}, people went beyond the cosmological constant and proposed several dynamical DE candidates/parametrizations. The evolution of any such beyond-CDM perturbations in these different DE backgrounds needs to be studied within proper theoretical frameworks. This is particularly important in the era of precision cosmology that we are entering into. With the advancements expected in forthcoming galaxy surveys \cite{Euclid:2011,Levi13,LSST:2017ags} even more accurate theoretical predictions are necessary. In particular, this will be important in probing the neutrino properties, their effects on perturbations and any possible information on the background physics. Theoretical calculations for precise observables like the cosmic microwave background (CMB) and the matter power spectrum, particularly for massive neutrinos, must meet stringent requirements. Neutrino masses are constrained by neutrino oscillation \cite{LSND:2001aii} and $\beta$-decay experiments \cite{KATRIN:2021uub} to be $\sum m_{\nu} \gtrsim 0.06$ eV, while CMB measurements \cite{Planck:2018vyg} and Baryon Acoustic Oscillation (BAO) data \cite{BOSS:2016wmc,Tanseri:2022zfe} constrain the sum of neutrino masses to $\sum m_{\nu} \lesssim 0.12$ eV. It is expected that these constraints will improve further in the light of future surveys such as Euclid \cite{Euclid:2011}, and LSST \cite{LSST:2017ags} which are aimed at providing more precise measurements about the recent universe and hence on the impact of massive neutrinos on cosmological structure formation. So, at this juncture any possible effects of massive neutrinos on cosmological evolution and perturbations need to be understood as clearly and precisely as possible, that may further reflect on the upcoming cosmological surveys.  

There are several approaches to understand the impact of massive neutrinos on the Large Scale Structure (LSS) formation of our universe. A robust approach is N-body simulation \cite{Ruggeri:2017dda,Bayer:2020tko} but they are rather computationally expensive particularly with massive neutrinos. Another popular approach is to solve the full Boltzmann hierarchy as implemented in the numerical code \texttt{CLASS} \cite{2011JCAP...09..032L} and  subsequently on the MCMC code \texttt{MontePython} \cite{Montepython}. From both the approaches it is well-established that massive neutrinos suppress the matter power spectrum because of free-streaming at small scales, while clustering like CDM at large scales. The approach which will be relevant for our study is to approximate massive neutrinos as a fluid with an effective sound speed. Such a fluid approximation, besides providing physical insights into structure formation, has been shown to be in good agreement with other frameworks \cite{Shoji:2010hm}. Such agreements have also resulted in mixed approaches combining the Boltzmann hierarchy and fluid approximation, matched at a certain redshift ($z$) \cite{Blas:2014hya}. In particular, the fluid approximation plays a crucial role in the computation of physical observables in the mildly nonlinear regime in Standard Perturbation Theory (SPT) \cite{Bernardeau:2001qr,Blas:2013aba}, Renormalised Perturbation Theory (RPT) \cite{Somogyi:2009mh,2014PhRvD..89b3502B}, Effective Field Theory (EFT) of LSS \cite{Carrasco:2012cv,Carrasco:2013mua,McQuinn:2015tva}, etc. When considering the impact of massive neutrinos on the  growth rate of structures in the universe \cite{Hu:1997mj,lesgourgues_mangano_miele_pastor_2013,Dupuy:2013jaa}, it is important to account for their scale-dependent behavior due to free-streaming. Future surveys aim to measure neutrino masses that reach a non-relativistic distribution long before nonlinear corrections become significant \cite{lesgourgues_mangano_miele_pastor_2013}. So, a consistent theoretical framework for both the linear and mildly nonlinear regimes, incorporating the time and scale dependent free-streaming behavior of massive neutrinos \cite{Blas:2014hya}, to be at par with future surveys, is of utmost priority today. 
In this article, we investigate a two-fluid setup consisting of CDM and massive neutrinos, coupled via gravitational interaction, where both the components are perturbing simultaneously in the expanding background. In this framework, we focus on the linear regime of structure formation ($k_{\rm NR} < k < k_{\rm NL}$) where $k_{\rm NR}$ is the neutrino non-relativistic scale and $k_{\rm NL}\sim 0.1 \mathrm{h} \; \mathrm{Mpc}^{-1}$ is the scale at which nonlinear effects become important. Besides being analytically tractable, the linear regime provides the kernels that are also useful while addressing the mildly nonlinear regime ( see \cite{Blas:2014hya,Saito:2008bp,Wong:2008ws} and \cite{Garny:2020ilv,Garny:2022fsh,Senatore:2017hyk} for some recent developments ). Moreover, adding non-gravitational neutrino interaction with dark matter in such a linearized fluid model framework is also an area of exciting phenomenology \cite{Green:2021gdc,Paul:2021ewd}.

Let us mention some of the salient features of our analysis. The essence of the two-fluid framework explored in this paper is to couple the neutrino fluid perturbations with the  perturbations of CDM (+ baryon) fluid via gravitational interactions.  This is significantly different from the single fluid framework \cite{1999} in which only the CDM sector is perturbed. In this two fluid framework, we work with physically realistic time-dependent neutrino free-streaming length {$k_{\rm FS}$} in the same vein as some of the previous works \cite{Shoji:2010hm,Wong:2008ws,2009ApJ...700..705S}. Observations suggest that DE starts to dominate at redshift $z \sim 0.4$, thus it is also important to incorporate the effects of various DE backgrounds while analyzing structure formation at late times. To this end, we perform our study incorporating the effect of $\Lambda$/DE in the fluid equations. In addition, we also extend our two-fluid formalism to  more general redshift-dependent DE backgrounds represented by different parametrizations. Finding a fully analytic solution to the coupled two-fluid equations including the aforementioned generalizations is somewhat difficult. However one can still make progress by treating the neutrino mass fraction ($f_{\nu}$) as a small parameter, a strategy that was advocated recently in \cite{Kamalinejad:2022yyl} and has also been followed in the present work. In this approach, we provide solutions in terms of  numerical integrals for the coupled two-fluid system in $\Lambda$CDM background and compute the resulting suppression in the total matter and velocity power spectrum. A separate approach towards finding an approximate fully analytic solution (without involving numerical integrals) is also presented for comparison and completeness.
For all the scenarios investigated in this work, we have compared the analytic results (arising from a  perturbative treatment in $f_{\nu}$) with numerical solutions of the two fluid equations (valid to all orders in $f_{\nu}$). The results obtained from both of these approaches are shown to be in agreement with numerical outputs of \texttt{CLASS} \cite{2011JCAP...09..032L} with at least $95 \%$ accuracy depending on the neutrino mass fraction. 
 
We then employ the above framework to analyze and estimate the  suppression in different DE backgrounds in the two-fluid setup. Finally, we find that suppression of matter power spectrum in different cosmological backgrounds lies within the $1\sigma$ uncertainty of BOSS-like galaxy survey noise for $\sum{m_{\nu}} = 0.12$ eV (corresponding $f_\nu =0.009$) which is the allowed upper bound from Planck \cite{Planck:2018vyg} and BAO data  \cite{BOSS:2016wmc}. Thus, we expect that the analysis would be crucial once the future LSS missions (+ CMB and 21-cm missions) are up and running. 

The structure of the paper is as follows. In Section \ref{role of Massive Neutrinos in LSS}, we review the role of massive neutrinos in structure formation. This is followed by the presentation of the two-fluid equations with time-dependent free-streaming length of neutrinos in Section \ref{Two-fluid equations} where we solve the two-fluid equations in presence of $\Lambda$ in a perturbative framework via two separate approaches and compute the suppression in total matter power spectrum. This is followed by a detailed numerical study of the power spectrum for different DE backgrounds beyond $\Lambda$CDM in Section \ref{Two-fluid equations in Different DE Universe}. The prospects in observations have been discussed in brief in Section \ref{Making Contact With Observations}. Finally, we conclude with a summary in Section \ref{Discussions and Outlooks} and comment on future directions. 

\section{Free-streaming of massive neutrinos and role in LSS}
\label{role of Massive Neutrinos in LSS}

For the sake of completeness, let us begin with a brief review of some important aspects of massive neutrinos and their role in cosmology with possible constraints from different observations. This will help in developing the rest of the article consistently. Details about neutrino cosmology can be found in some excellent reviews \cite{lesgourgues_mangano_miele_pastor_2013,Lesgourgues:2006nd}. Solar and Atmospheric neutrino oscillation experiments \cite{2012JHEP...12..123G} constrain the neutrino mass squared differences to be,
\begin{align}
    \Delta m_{21}^2 = (7.50^{+0.18}_{-0.19})\times 10^{-5}\; {\rm{eV}}^2, \;\;\vert\Delta m_{31}^2 \vert=(2.473^{+0.042}_{-0.065})\times 10^{-3}\; \rm{eV}^2.
\end{align}
Along with that, cosmological constraints on neutrinos primarily come from two observations: CMB \cite{Planck:2018vyg} and LSS data \cite{BOSS:2016wmc}. Together they constrain the sum of neutrino masses to be, 
\begin{align}
    0.06\, \rm{eV} \lesssim \sum_{i} m_{\nu,i} \lesssim 0.12\,\rm{eV}.
\label{massrange} 
\end{align}

As far as current observational probes are concerned, cosmological large scale structure formation is insensitive to the neutrino mass hierarchy \cite{archidiacono2020will}. Thus it is sufficient to consider the effects of sum of neutrino mass with degenerate mass states. Neutrinos contribute as a radiation component in the early universe while they behave like matter in the late universe. In the early universe, neutrinos decouple from the primordial plasma when the temperature of the universe is $T \approx 1\; \rm{MeV}$ and thereafter free-stream as a massless relic. With the expansion of the universe the momentum of the neutrinos falls as $1/a$ where $a$ is the scale factor and hence they become non-relativistic at a certain redshift. The comoving energy, $\epsilon_i$, of the  $i^{th}$ neutrino mass state is related to the comoving momentum ($\tilde{q}$) and rest mass of the neutrinos through the relation $\epsilon_i = \sqrt{\tilde{q}_i^2+a^2m_{\nu,i}^2}$. In the relativistic limit, the average momentum of the neutrinos can be obtained from the Fermi-Dirac distribution \cite{Shoji:2010hm,Blas:2014hya,Levi2016tlf} as, 
\begin{eqnarray}
    \left<p\right>&=&\frac{\int d^3\mathbf{p}\;p\, \left[e^{\left(p/T_{\nu}\right)}+1\right]^{-1}}{\int d^3\mathbf{p}\, \left[e^{\left(p/T_{\nu}\right)}+1\right]^{-1}}\nonumber \\
    &=&3.15\; T_{\nu,0}\;(1+z),
\label{mean momentum}
\end{eqnarray}
where $T_{\nu,0} \left[= \left(4/11 \right)^{1/3}T_{\rm CMB}\right]$ is the neutrino temperature at present and $T_{\rm CMB}$ is the CMB temperature. Therefore the redshift of non-relativistic transition of the $i^{th}$ neutrino mass state, $z_{\mathrm{NR},i}$, can be obtained from the relation,
\begin{align}
    1+z_{\mathrm{NR},i} = 1890 \left(\frac{m_{\nu,i}}{1\;\rm{eV}}\right).
\label{transition redshift}
\end{align}
Once the neutrinos are non-relativistic, the ratio of the neutrino mass density to the total matter density $(\frac{\Omega_\nu}{\Omega_{\rm m}})$ denoted by $f_\nu$ becomes constant where the mass density in the neutrino sector can be obtained from
\begin{align}
    \Omega_\nu h^2 = \frac{\sum_{i} m_{\nu,i}}{93.14\,eV}.
\end{align}

Neutrino density fluctuations cannot grow within the horizon until the non-relativistic transition, as large thermal velocity of neutrinos impedes the gravitational instability to grow. Thus non-relativistic transition imprints a characteristic scale to the evolution of neutrino density fluctuations. The scale above which density fluctuations in neutrino sector begins to grow is commonly referred as the free-streaming length. The evolution of velocity dispersion with redshift can be obtained from the phase space distribution function of neutrinos following,
\begin{eqnarray}
    {\sigma^2_{\nu,i}}(z) \equiv\frac{\int d^3\mathbf{p}\;\frac{p^2}{m^2_{\nu,i}}\, \left[e^{\left(p/T_{\nu}\right)}+1\right]^{-1}}{\int d^3\mathbf{p}\, \left[e^{\left(p/T_{\nu}\right)}+1\right]^{-1}}
    = \frac{15\;\zeta(5)}{\zeta(3)}\;\frac{T^2_{\nu,0}\;(1+z)^2}{m^2_\nu,i}.
\label{velocity dispersion}
\end{eqnarray}
On the other hand the free-streaming length can be defined in terms of neutrino sound speed $ c^2_{\mathrm{s}}(z) $, \footnote{The sound speed can be obtained from the relation $ c^2_{\mathrm{s}}(z)\equiv \frac{\delta P}{\delta \rho} $, starting from neutrino phase space distribution. Here we have used the approximation $c^2_{\mathrm{s}}=\frac{5}{9}\sigma^2_{\nu,i}$ following \cite{Shoji:2010hm} } as,
\begin{align}
    {{ k^2_{\rm FS}}(z)} \equiv \frac{3}{2}\Omega_{\rm m}(z) \frac{\mathcal{H}^2(z)}{ c^2_{\mathrm{s}}(z)} = \frac{27}{10}\Omega_{\rm m}(z) \frac{\mathcal{H}^2(z)}{{{ \sigma^2_{\nu}}}(z)},
\label{kfs}
\end{align}
where $\Omega_{\rm m}(z)$ is the fractional matter density at redshift $z$. As mentioned earlier, due to large thermal velocity, neutrinos do not cluster on scales smaller than $k_{\rm FS}$, while on larger scales they behave like cold dark matter (CDM) \cite{Levi16}. So one can approximate the sound speed of the neutrino fluid and hence the free-streaming length in terms of the velocity dispersion of the non-relativistic neutrinos \cite{Shoji:2010hm}. Combining eqs. (\ref{velocity dispersion}) and (\ref{kfs}), the free-streaming length $(k_{\rm FS})$ at sufficiently late time $a \gg a_{\mathrm{NR}}$ \cite{Slepian_2018_neutrino, Shoji:2010hm} can be approximated as,
\begin{align}
    {k^2_{\rm FS}}(z) = {k^2_0} \;a(z) 
\label{kfree}
\end{align}
 where $a_{\mathrm{NR}}$ is the scale factor at the time of non-relativistic transition and $k_0$ is the neutrino free-streaming length today,
\begin{align}
    k_0 = \left(\frac{9\:\zeta(3)}{50\;\zeta(5)}\frac{\;H_0^2\;\Omega_{{\rm m}, 0}\;m^2_{\nu,i}}{{T^2_{\nu,0}}}\right)^{\frac{1}{2}}. 
\label{kfree0}
\end{align}
The free-streaming scale $k_{\rm FS}$ first decreases with time and reaches a minimum around the non-relativistic transition epoch defined as $k_{\rm NR}$ after that it increases with time like eq.~(\ref{kfree}).
The validity of the approximation has been explored in detail in \cite{Shoji:2010hm,Kamalinejad:2022yyl} at late time, particularly in matter dominated era. Here we have analysed the validity of the approximation in generalised background cosmology in a perturbative fashion in the followed sections.

 \section{Coupled CDM-massive neutrino perturbations in \texorpdfstring{$\Lambda$}{}CDM background}
 \label{Two-fluid equations}

In this section, we build up all the necessary equations and definitions needed for our theoretical framework in the conformal-Newtonian gauge.
The equations for the two-fluid model of CDM and massive neutrinos are obtained by taking moments of the Boltzmann equation with the perturbed phase space distribution function. 
The perturbed phase space distribution for neutrinos upto linear order is defined as,
\begin{align}
    f(\mathbf{k}\cdot\mathbf{\hat{n}},\tilde{q},\eta)\;=\;f_0(\tilde{q},\eta)\;[1\;+\;\Psi(\mathbf{k}\cdot\mathbf{\hat{n}},\tilde{q},\eta)]
\end{align}
where $\Psi(\mathbf{k}\cdot\mathbf{\hat{n}},\tilde{q},\eta)$ is the linear order perturbation, $ \tilde{q} \equiv \vert \tilde{q} \vert$ is the magnitude of the comoving momentum and $\hat{n} \equiv \mathbf{\tilde{q}}/\tilde{q}$.
$f_0(\tilde{q},\eta)$ is the relativistic Fermi-Dirac distribution given by,
\begin{align}
    f_0(\tilde{q})\;=\;\frac{g_s}{(2\pi)^3}\frac{1}{e^{(\tilde{q}/T_{\nu,0})}+1}
\end{align}
where $T_{\nu,0} \approx 1.95 \mathrm{K} \approx 1.7 \times 10^{-4}\;eV$ is the neutrino temperature today and $g_s$ is the degeneracy factor of neutrinos.\\
Following \cite{Ma:1995ey,Shoji:2010hm} the evolution of linear order perturbation for collision-less massive neutrinos is given by the Boltzmann equation,
\begin{eqnarray}
\frac{\partial\Psi(\mathbf{k}\cdot\mathbf{\hat{n}},\tilde{q},\eta)}
{\partial\eta}+i\frac{\tilde{q}}{\epsilon(\tilde{q},\eta)}
\mu\Psi(\mathbf{k}\cdot\mathbf{\hat{n}},\tilde{q},\eta)+\frac{d\ln f_0(\tilde{q})}{d\ln \tilde{q}}\left(
\dot{\phi}(k,\eta)-i\frac{\epsilon(\tilde{q},\eta)}{\tilde{q}} \mu \psi(k,\eta)
\right)=0, 
\end{eqnarray}   
where $\mathbf{k}\cdot\mathbf{\hat{n}} = k \mu$ is the angle between the wavenumber and the comoving momentum, $\epsilon(\tilde{q},\eta)$ is the comoving energy, $\psi(k,\eta)$ and $\phi(k,\eta)$ are the gravitational potentials.

The standard procedure to deal with the equation is to expand $\Psi(\mathbf{k}\cdot\mathbf{\hat{n}},\tilde{q},\eta)$ in multipoles and we get an infinite hierarchy of equations for multipoles. When considering the non-relativistic limit, the resulting hierarchy of equations can be truncated at $\ell=1$, which results in a set of coupled differential equations that describes the evolution of the density contrast $\delta$ and velocity divergence $\theta$. This two-fluid treatment introduces a distinct free-streaming length scale, similar to the Jeans scale, into the fluid equations due to the presence of massive neutrinos. The validity of the fluid approximation including massive neutrinos has been extensively studied in \cite{Shoji:2010hm,Blas:2014hya}, and recently in \cite{Nascimento:2023psl}. In the non-relativistic limit, the coupled fluid equations are given as follows (ignoring vorticity and nonlinear terms),
\begin{align}
    &\dot{\delta}_{\rm cb}+\theta_{\rm cb}-3\dot{\phi}(k,\eta)=0,\label{dcb}\\
    &\dot{\theta}_{\rm cb}+\mathcal{H}\theta_{\rm cb}+\frac{3}{2}\mathcal{H}^{2}\Omega_{\rm m}(\eta)\left( f_{\nu}\delta_{\nu}+(1-f_{\nu})\delta_{\rm cb} \right) = 0,\label{tcb}\\
    &\dot{\delta}_{\nu}+\theta_{\nu} -3\dot{\phi}(k,\eta)= 0,\label{dnu}\\
    &\dot{\theta}_{\nu}+\mathcal{H}\theta_{\nu}+\frac{3}{2}\mathcal{H}^{2}\Omega_{\rm m}(\eta)\left(f_{\nu}\delta_{\nu}+(1-f_{\nu})\delta_{\rm cb} \right) -k^2 {c_s}^2 \delta_{\nu} = 0,
    \label{tnu}
\end{align}
where $\mathcal{H}$ is the Hubble parameter in terms of conformal time $\eta$ with $d\eta \equiv dt/a(t)$ and the derivatives are also with respect to conformal time. Also, the suffix ``${\rm cb}$'' represents combined CDM+baryon, considered as a single fluid in the present analysis and identified as ``CDM'' henceforth, and ``$\nu$'' represents massive neutrinos. $\phi(k,\eta)$ appearing in the above set of equations is the gravitational potential in Newtonian gauge \cite{Shoji:2010hm,Ma:1995ey} and is coupled to the total matter density perturbations via the Poisson equation, 
\begin{align}
    k^2 \phi(k,\eta) = \frac{3}{2}\mathcal{H}^2(\eta)~\Omega_{\rm m}(\eta)~\delta_{\rm{m}}(k,\eta).
    \label{Poisson}
\end{align}
where $\delta_{\rm{m}}(k,\eta)$ \footnote{$\eta$ represents conformal time whereas $\tau$ denotes $\ln(a)$ throughout the article}is the total matter overdensity defined as,
\begin{eqnarray}
    \delta_{\rm{m}}(k,\eta) = f_{\nu} \delta_{\nu}(k,\eta) + \left(1-f_{\nu}\right) \delta_{\rm{cb}}(k,\eta).
\label{matterDP}
\end{eqnarray} 

Having the stage set up, we proceed with our analysis to solve the coupled fluid equations in $\Lambda$CDM universe in the following subsection. The evolution of neutrino overdensity and velocity divergence in matter-only Einstein-de Sitter (EDS) universe has been explored to a considerable extent in \cite{Kamalinejad:2022yyl, lesgourgues_mangano_miele_pastor_2013}. However, in order to corroborate with observations, one needs to go beyond a matter-only Universe and develop a consistent framework in background  $\Lambda$CDM (and possibly, evolving DE) cosmology.

Specifically, our primary target is to obtain a consistent set of solutions of the coupled two fluid system considering the Hubble parameter $\mathcal{H}$ appropriate for $\Lambda$CDM evolution. Combining eqs.~\eqref{dcb}-\eqref{tnu} we obtain the following two coupled second-order differential equations for CDM and neutrino fluids, 
\begin{eqnarray}
    &{\delta}_{\rm{cb}}^{''}+\left(1+\frac{{\mathcal{H}}^{'}}{\mathcal{H}}\right){\delta}_{\rm{cb}}^{'} = 3 \phi^{''}+3 \left(1+\frac{{\mathcal{H}}^{'}}{\mathcal{H}}\right) \phi^{
'} +\frac{3}{2}\Omega_{\rm{m}}\left( f_{\nu}\delta_{\nu}+(1-f_{\nu}){\delta}_{\rm{cb}} \right),\label{dcb''}\\
    & \delta_{\nu}^{''}+\left(1+\frac{{\mathcal{H}}^{'}}{\mathcal{H}}\right)\delta_{\nu}^{'} = 3 \phi^{''}+3 \left(1+\frac{{\mathcal{H}}^{'}}{\mathcal{H}}\right) \phi^{
'} +\frac{3}{2}\Omega_{\rm{m}}\left(f_{\nu}  \delta_{\nu}+(1-f_{\nu}) {\delta}_{\rm{cb}}-\left(\frac{k}{k_{\rm{FS}}}\right)^{2} \delta_{\nu} \right) \label{dnu''}
\end{eqnarray}
where the derivatives are with respect to  $\tau \equiv \ln{a}$. Since we are primarily interested in an epoch where neutrinos have long become non-relativistic, we have considered $f_\nu$ to be time independent \cite{Shoji:2010hm}. However the free-streaming scale is time-dependent through eq.~\eqref{kfree}. 
Note that in  $\Lambda$CDM background we have,
\begin{align}
1+\frac{\mathcal{H}^{'}}{\mathcal{H}} = 2-\frac{3}{2(1+\tilde{\Lambda}  e^{3 \tau})}, ~\Omega_m(\tau) =\frac{1}{(1+\tilde{\Lambda}  e^{3 \tau})}
\end{align}
where $\tilde{\Lambda}\equiv \frac{\Omega_{\Lambda,0}}{\Omega_{\rm m,0}}$ where $\Omega_{\Lambda,0}$ and $\Omega_{\rm m,0}$ are respectively DE and matter density today. We will further keep in mind, that unlike EDS, $\Omega_{\rm m,0} \sim 0.3$ and $\Omega_{\Lambda,0} \sim 0.7$ for $\Lambda$CDM.

One can readily identify that the above set of equations can in principle represent both EDS and $\Lambda$CDM universe, for particular values of the parameter $\tilde{\Lambda}$. However, due to time-evolution associated with this parameter, the background evolution and hence the above set of perturbation equations are significantly different from EDS. Our analysis thus extends the work done in \cite{Kamalinejad:2022yyl} from theoretical perspectives with distinctive features of the realistic observable universe. As remarked earlier, we will mostly focus on the $\Lambda$CDM (and beyond) case in the subsequent analysis.

\subsection{Perturbative expansion in \texorpdfstring{$f_\nu$}{}}
\label{Perturbative expansion}

We now engage ourselves in finding  a consistent set of solutions to eqs.~\eqref{dcb''} and \eqref{dnu''}. To this end, we will do a perturbative expansion of the above set of equations up to first order in $f_\nu$ in the $\Lambda$CDM background. Following standard prescription, the CDM and neutrino density constrasts can be expressed in a perturbative series as,
\begin{eqnarray}
    \delta_{\rm{cb}} &= \delta_{\rm{cb}}^{(0)}+\delta_{\rm{cb}}^{(1)}+\cdots,\\
\label{dcbexpand}
    \delta_{\nu} &= \delta_{\nu}^{(0)}+\delta_{\nu}^{(1)}+\cdots
\label{dnuexpand}
\end{eqnarray}
where $\delta_i^{(n)}$ is the $n^{th}$ order perturbation in $f_\nu$ of the $i^{th}$ species. Also it is evident from eq.~\eqref{Poisson} that at late times, the gravitational potential $\phi$ decays as $\frac{1}{k^2}$ compared to other terms, hence we neglect any contribution from $\phi$ in eqs.~\eqref{dcb''} and \eqref{dnu''} for simplicity. As a result, for $a\gg a_{\mathrm{NR}}$, the zeroth-order density perturbation equations for CDM and neutrino sectors take the form,
\begin{eqnarray}
    & {\delta_{\rm cb}^{(0)}}^{\prime\prime}+\left(2-\frac{3}{2(1+\tilde{\Lambda} e^{3\tau})}\right){\delta_{\rm cb}^{(0)}}^{\prime} -\frac{3}{2(1+\tilde{\Lambda} e^{3\tau})}\delta_{\rm cb}^{(0)}=0 \label{dcb0order},\\
\nonumber\\
    &{\delta_{\nu}^{(0)}}^{\prime\prime} +\left(2-\frac{3}{2(1+\tilde{\Lambda}e^{3\tau})}\right){\delta_{\nu}^{(0)}}^{\prime} +\frac{3}{2}q^2\frac{e^{-\tau}}{(1+\tilde{\Lambda} e^{3\tau})}\delta_{\nu}^{(0)}-\frac{3}{2(1+\tilde{\Lambda} e^{3\tau})}\delta_{cb}^{(0)}=0
\label{dnu0order}
\end{eqnarray}
where $q \equiv k/k_0$ ($k_0$ is defined in eq.~\eqref{kfree0}). We remind the reader that throughout this article we consider CDM+baryon as single fluid ``cb'' and call it ``CDM'' for brevity. Likewise, the first-order density perturbation equations for CDM and neutrinos turn out to be,
\begin{eqnarray}
    {\delta_{\rm cb}^{(1)}}^{\prime\prime} +\left(2-\frac{3}{2(1+\tilde{\Lambda} e^{3\tau})}\right){\delta_{\rm cb}^{(1)}}^{\prime} -\frac{3}{2(1+\tilde{\Lambda} e^{3\tau})}\delta_{\rm cb}^{(1)}+\frac{3}{2(1+\tilde{\Lambda}e^{3\tau})}f_{\nu}({\delta_{\rm cb}^{(0)}}-{\delta_{\nu}^{(0)}})=0
    \label{dcb1order}
\end{eqnarray}
\begin{align}
    {\delta_{\nu}^{(1)}}^{\prime\prime}+(2-\frac{3}{2(1+\tilde{\Lambda} e^{3\tau})}){\delta_{\rm cb}^{(1)}}^{\prime}+\frac{3}{2}q^2\frac{e^{-\tau}}{(1+\tilde{\Lambda} e^{3\tau})}\delta_{\nu}^{(1)}-\frac{3}{2(1+\tilde{\Lambda} e^{3\tau})}\delta_{\rm cb}^{(1)}+\frac{3}{2(1+\tilde{\Lambda}e^{3\tau})}f_{\nu}{\delta_{\rm cb}^{(0)}}=0
\label{dnu1order}
\end{align}
The eqs.~\eqref{dcb0order}-\eqref{dcb1order} show that the CDM perturbations of zeroth order will source the neutrino perturbations of zeroth order, which in turn sources the CDM perturbations at first order as indicated in eq.~(\ref{dcb1order}). Since our focus is on getting the solutions at the lowest order in $f_{\nu}$, we only need to solve the zeroth-order equations for each species and then solve the first-order equation for CDM. We will present the solutions of the above equations in the next subsections following two different methods. In the first approach, we find exact solutions to the perturbation  eqs.~\eqref{dcb0order}-\eqref{dcb1order} in terms of numerical integrals.  In the second approach, we first solve for the growth factor $\ln D(a)$ of the CDM incorporating the effects of the background in the absence of neutrinos and then solve the perturbative equations following ``EDS approximation''. Further, in Section \ref{Two-fluid equations in Different DE Universe}, we will expand the scope of this framework to investigate the impact of neutrino perturbations on the matter power spectrum in various DE models.

\subsection{Exact solution to perturbation equations in $\Lambda$CDM universe}
\label{solutions}

In order to find out a consistent set of solutions to the above set of perturbation equations we will first exploit the zeroth order eq.~\eqref{dcb0order} for CDM density contrast. Ignoring the decaying mode, the physically relevant solution for $\delta_{\rm{cb}}^{(0)}$ is given by,
\begin{align}
\delta_{\rm cb}^{(0)}(q,\tau)= c_0(q) e^{\tau}\sqrt{1+\tilde{\Lambda} e^{3\tau}}\;{}_2F_1 \left(\frac{5}{6},\frac{3}{2},\frac{11}{6},-\tilde{\Lambda} e^{3\tau}\right) ,
\label{0orderdcb}
\end{align}
where ${}_2F_1$ denotes the Gauss-Hypergeometric function and the integration constant $c_0(q)$ can be redefined either in terms of $\delta_{\rm{cb}}(q,0)$ which is the CDM density contrast today in absence of neutrinos or in terms of CDM density contrast at the time of neutrino non-relativistic transition.

Next we will explore the zeroth order neutrino density perturbation eq.~\eqref{dnu0order}. As can be seen, this equation contains inputs from zeroth order CDM perturbations, the solution of which is given by eq.~(\ref{0orderdcb}). Inserting this CDM density contrast into eq.~(\ref{dnu0order}), the zeroth order neutrino density contrast turns out to be,

\begin{align} 
 & \delta_{\nu}^{(0)}(q,\tau)= c_1 (q)\cos \left[\sqrt{6}q e^{-{\frac{\tau}{2}}}{}_2F_1^1(\tau)\right]-c_2(q)\sin \left[\sqrt{6}q e^{-{\frac{\tau}{2}}}{}_2F_1^1(\tau)\right]+ \nonumber \\& \cos \left[\sqrt{6}q e^{-{\frac{\tau}{2}}}{}_2F_1^1(\tau)\right] \int_{1}^{\tau} \frac{1}{q}\sqrt{\frac{3}{2}}e^{\frac{3\tau^{\prime}}{2}}{}_2F_1^2(\tau^\prime) c_0 (q)\sin \left[\sqrt{6}q e^{-{\frac{\tau^{\prime}}{2}}}{}_2F_1^1(\tau^{\prime})\right] \,d\tau^{\prime}- \nonumber \\& \sin \left[\sqrt{6}q e^{-{\frac{\tau}{2}}}{}_2F_1^1(\tau)\right] \int_{1}^{\tau} \frac{1}{q}\sqrt{\frac{3}{2}}e^{\frac{3\tau^{\prime}}{2}}{}_2F_1^2(\tau^\prime) c_0 (q) \cos \left[\sqrt{6}q e^{-{\frac{\tau^{\prime}}{2}}}{}_2F_1^1(\tau^{\prime})\right]\,d\tau^{\prime}
 \label{dnu00order}
 \end{align}
where $c_1(q)$ and $c_2(q)$ can be related to any suitable initial condition for the neutrino perturbation at the transition redshift. For brevity here, and throughout the rest of the article, we will denote by ${}_2F_1^1(\tau)$ and ${}_2F_1^2(\tau)$ respectively, the following two specific Gauss-Hypergeometric functions,

\begin{align}
    {}_2F_1^1(\tau) \equiv {}_2F_1\left(-\frac{1}{6},\frac{1}{2},\frac{5}{6},-\tilde{\Lambda}\, e^{3\tau}\right) \nonumber \\
    {}_2F_1^2(\tau) \equiv {}_2F_1\left (\frac{5}{6},\frac{3}{2},\frac{11}{6},-\tilde{\Lambda} \,e^{3\tau}\right).
    \label{hypergnota}
\end{align}

Finally, we need to find out the solution for eq.~(\ref{dcb1order}) in order to compute the matter power spectrum up to first order in $f_\nu$. Using the solution of $\delta_{\rm{cb}}^{(0)}$ and $\delta_{\nu}^{(0)}$ from eqs.~\eqref{0orderdcb} and \eqref{dnu00order} we obtain the following solution for the first order CDM density contrast,
\begin{align}
&\delta_{\rm{cb}}^{(1)}(q,\tau) =  c_4(q)\,e^{\tau}\sqrt{1+\tilde{\Lambda}e^{3\tau}}\,{}_2F_1^2(\tau) \int_{0}^{\tau}  \frac{18}{25}\,f_{\nu}\, e^{\frac{3}{2}\tau^{\prime}}{}_2F_1^1(\tau^{\prime}) \left(\delta_{\rm cb}^{(0)}(q,\tau^{\prime})-\delta_{\nu}^{(0)}(q,\tau^{\prime})\right)\,d\tau^{\prime}+ \nonumber \\&   c_3(q)\,e^{\frac{-3\tau}{2}}\sqrt{1+\tilde{\Lambda}\;e^{3\tau}} + e^{\tau}\sqrt{1+\tilde{\Lambda}\;e^{3\tau}}\,{}_2F_1^1(\tau) \int_{0}^{\tau}\frac{3}{5}f_{\nu}e^{-\tau^{\prime}} \left(\delta_{\nu}^{(0)}(q,\tau^{\prime})-\delta_{\rm cb}^{(0)}(q,\tau^{\prime})\right) \,d\tau^{\prime}
\label{dcbwi1}
\end{align}
We will utilize the above solutions with reasonable initial conditions to compute the power spectrum correct up to first order in $f_\nu$ in Section.~\ref{suppression in the power spectra}.

\subsection{Approximate solution to  perturbation equations in $\Lambda$CDM universe: Alternate approach}
\label{method2}
In this section we present an alternate approach to arrive at an approximate but fully analytic (i.e. without involving numerical integrals) solution of the coupled two fluid system. We first rewrite eqs.~(\ref{dcb''}) and (\ref{dnu''}) in terms of a new variable $ s=\ln(D(a))$ where $D(a)$ is the growth factor in the absence of neutrinos. Upon ignoring terms involving  $\phi^\prime$'s at late times as explained earlier, we arrive at the following equations, 

\begin{eqnarray}
   &\delta_\mathrm{cb}^{\prime \prime}(s)+ \left(\frac{3}{2}\;\frac{\Omega_{\mathrm{m}}}{f^2}-1\right)\delta_\mathrm{cb}^{\prime}(s) - \frac{3}{2}\; \frac{\Omega_{\mathrm{m}}}{f^2}\;\left(f_\nu \delta_\nu(s) +(1-f_\nu)\delta_{\mathrm{cb}}(s)\right)=0 
   \label{dcb2''}
   \\
   &\delta_\nu^{\prime \prime}(s)+ \left(\frac{3}{2}\;\frac{\Omega_{\mathrm{m}}}{f^2}-1\right)\delta_\nu^{\prime}(s) -\frac{3}{2}\; \frac{\Omega_{\mathrm{m}}}{f^2}\;\left(f_\nu \delta_\nu(s) +(1-f_\nu)\delta_{\mathrm{cb}}(s)-\frac{k^2}{k^2_{\mathrm{FS}}}\delta_\nu(s) \right)=0 
   \label{dnu2''}
\end{eqnarray}
where the derivatives are w.r.t. the new variable $s$. $\Omega_{\rm{m}}$ is the fractional matter density as before and $f$ is the growth rate\footnote{The growth rate is defined as $f = \frac{d\ln(D(a))}{d\ln(a)}$ where $D(a)$ is the growth factor as mentioned.} which is not to be confused with the neutrino mass fraction $f_\nu$. The term $\frac{\Omega_{\rm{m}}}{f^2}$ encapsulates the effect of $\Lambda$ in $\Lambda$CDM cosmology in the above set of equations. To leading order we can approximate the ratio to be $1$ \cite{Bernardeau:2001qr, Garny:2020ilv}, which is different from setting $\Omega_{\rm{m}}=1$ and $f=1$ individually as is usually done in EDS universe. The effect of $\Lambda$ in this approach resides in the growth factor $D(a)$. The key advantage of using this approximation is that now the perturbation equations obtained from eqs.~(\ref{dcb2''}) and (\ref{dnu2''}) are identical to the equations relevant for EDS universe \cite{Kamalinejad:2022yyl}, except that now the new time variable $s$ incorporates the effects from $\Lambda$. To solve for $\delta_{\nu}^{(0)}$ and $\delta_{\rm{cb}}^{(1)}$ in eqs.~(\ref{dcb2''}) and (\ref{dnu2''}) (following the perturbative prescription as before), we simply recast the solutions of \cite{Kamalinejad:2022yyl} in terms of the growth factor $s(\tau)=\ln(D(\tau))=\ln \left[e^{\tau}\sqrt{1+\tilde{\Lambda} \,e^{3\tau}}\;{}_2F_1 (\frac{5}{6},\frac{3}{2},\frac{11}{6},-\tilde{\Lambda}\, e^{3\tau}) \right]$ . In this way we get a full scale dependent solution for $\delta_{\nu}^{(0)}$ in terms of $\tau$ and nonzero $\Lambda$ as presented below,
\begin{align}
&\delta_{\nu}^{(0)}(q,\tau) = c_0(q)\, e^{\tau } \sqrt{1+\tilde{\Lambda} \, e^{3 \tau }} \, {}_2F_1\left (\frac{5}{6},\frac{3}{2},\frac{11}{6},-\tilde{\Lambda} \,e^{3\tau}\right)\nonumber \\& + c_1(q)\, \cos \left[\sqrt{6} q {\left(e^{\tau } \sqrt{1+\tilde{\Lambda} \, e^{3 \tau }} \, {}_2F_1^2(\tau)\right)}^{-\frac{1}{2}}\right]-c_2(q)\, \sin \left[\sqrt{6} q \left(e^{\tau } \sqrt{1+\tilde{\Lambda}\, e^{3 \tau }} \, {}_2F_1^2(\tau)\right){}^{-\frac{1}{2}}\right]\nonumber \\& +  c_0(q)\, 6 q^2 \,\cos \left[\sqrt{6} q {\left(e^{\tau } \sqrt{1+\tilde{\Lambda} \, e^{3 \tau }} \, {}_2F_1^2(\tau)\right)}^{-\frac{1}{2}}\right] \text{Ci}\left[\sqrt{6} q {\left(e^{\tau } \sqrt{1+\tilde{\Lambda} \, e^{3 \tau }} \, {}_2F_1^2(\tau)\right)}^{-\frac{1}{2}}\right] \nonumber \\& +c_0(q) \, 6 q^2 \,\sin \left[\sqrt{6} q {\left(e^{\tau } \sqrt{1+\tilde{\Lambda}\, e^{3 \tau }} \, {}_2F_1^2(\tau)\right)}^{-\frac{1}{2}}\right] \text{Si}\left[\sqrt{6} q {\left(e^{\tau } \sqrt{1+\tilde{\Lambda}\, e^{3 \tau }} \, {}_2F_1^2(\tau)\right)}^{-\frac{1}{2}}\right]\nonumber \\&
\label{deltanu0}
\end{align}
where $\text{Si}$ and $\text{Ci}$ respectively denote the Sin and Cosine integral functions. As before, the initial conditions are set in such a way that the neutrino perturbation traces CDM in large scale and vanishes at small scales. We have exploited the role of initial conditions in the following subsection \ref{transferfunction}. Also the equation for $\delta_{\rm{cb}}^{(1)}$ in eq.~(\ref{dcb2''}) is sourced by the neutrino perturbation and gives the solution in terms of a time integral. It takes the following form, 
\begin{align}
  &\delta_{cb}^{(1)}(q,\tau) = c_3(q)\, e^{\tau } \sqrt{1+\tilde{\Lambda} \,e^{3 \tau }} \, {}_2F_1^2(\tau) + c_4(q)\, {\left(e^{\tau } \sqrt{1+\tilde{\Lambda} \,e^{3 \tau }} \, {}_2F_1^2(\tau)\right)}^{-\frac{3}{2}} \nonumber \\& {\left(e^{\tau } \sqrt{1+\tilde{\Lambda} \,e^{3 \tau }} \, {}_2F_1^2(\tau)\right)}^{-\frac{3}{2}} \int _1^{\ln \left(e^{\tau } \sqrt{1+\tilde{\Lambda} \, e^{3 \tau }} \, {}_2F_1^2(\tau)\right)}\frac{3}{5}f_{\nu} e^{\frac{3}{2}\tau^{\prime}}{}_2F_1^1(\tau^{\prime}) \left(\delta_{\rm cb}^{(0)}(q,\tau^{\prime})-\delta_{\nu}^{(0)}(q,\tau^{\prime})\right)\,d\tau^{\prime} \nonumber \\& +e^{\tau } \sqrt{1+\tilde{\Lambda} \, e^{3 \tau }} \, {}_2F_1^2(\tau) \int _1^{\ln \left(e^{\tau } \sqrt{1+\tilde{\Lambda} \, e^{3 \tau }} \, {}_2F_1^2(\tau)\right)}\frac{3}{5}f_{\nu}e^{-\tau^{\prime}} \left(\delta_{\nu}^{(0)}(q,\tau^{\prime})-\delta_{\rm cb}^{(0)}(q,\tau^{\prime})\right) \,d\tau^{\prime}
  \label{deltacb1order}
\end{align}
Combining the eqs.~(\ref{0orderdcb}), (\ref{deltanu0}) and (\ref{deltacb1order}), we obtain a fully analytic, scale- and time-dependent solution, from which we extract the suppression of the matter power spectra as before. This will be presented in figure \ref{fig:analytical}.

In the following subsections, we have performed a detailed comparison of the analytical solutions with the numerical solutions of the two fluid systems both for $\Lambda$CDM and $w$CDM cosmologies. Note that we will refer to the exact solution to the perturbation equations as ``Method 1'' and the approximate fully analytic solution as ``Method 2'' in the following subsections.

\subsection{Transfer function and role of initial conditions}
\label{transferfunction}

\begin{figure}[ht!]
    \centering
    \subfloat{\includegraphics[width= 0.35\textwidth]{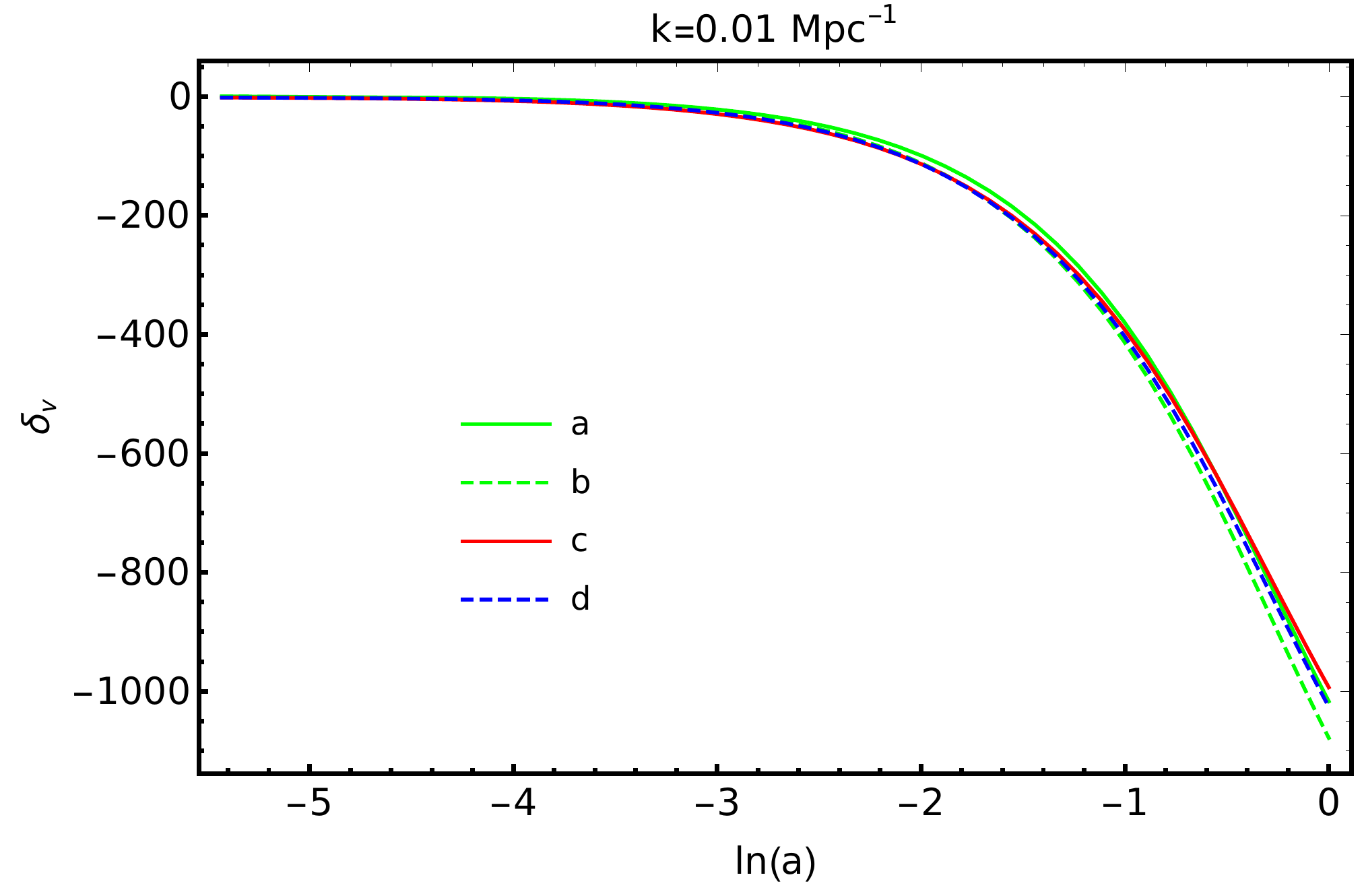} }
    \subfloat{\includegraphics[width=0.35\textwidth]{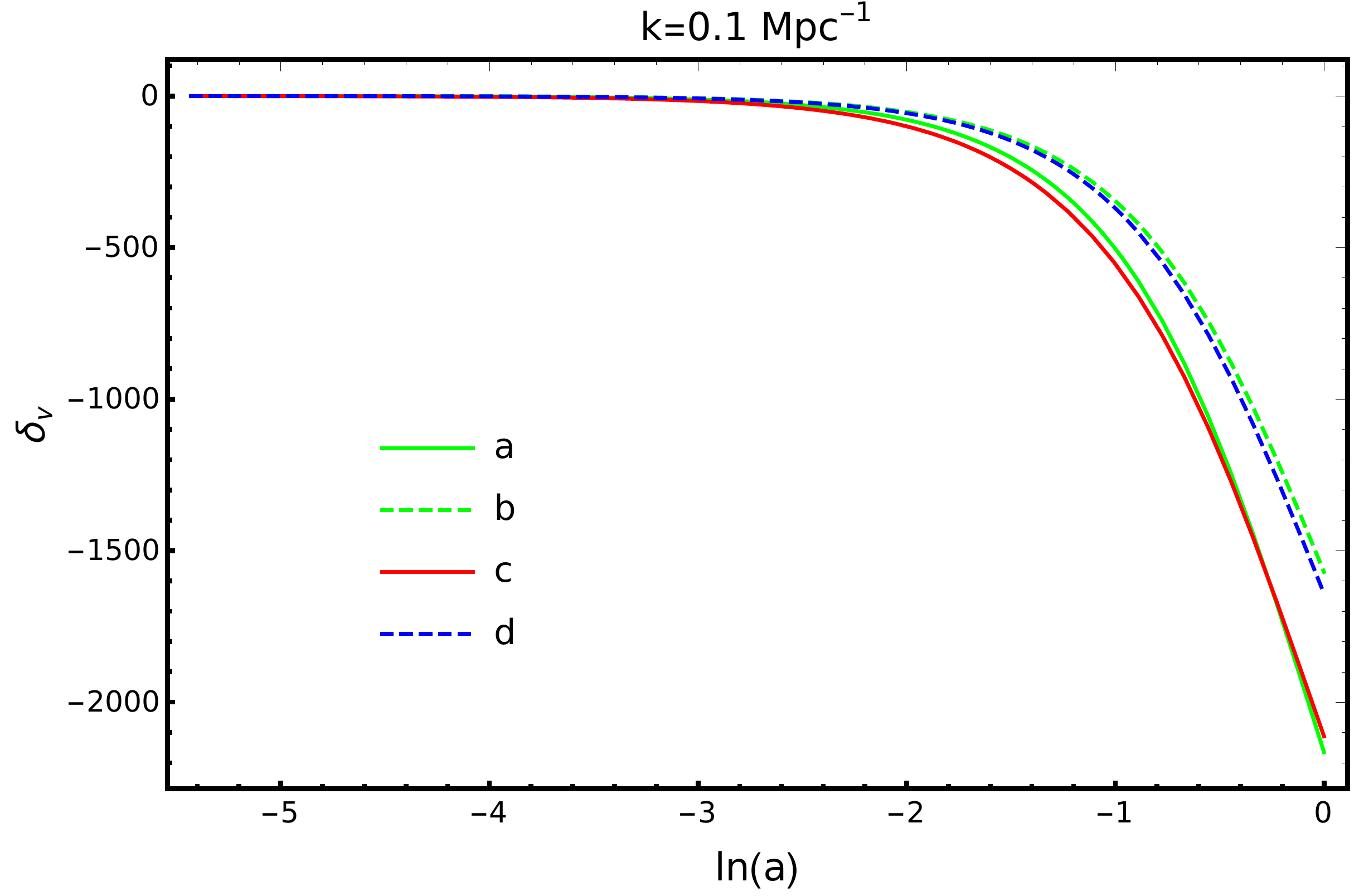} }
    \subfloat{\includegraphics[width=0.35\textwidth]{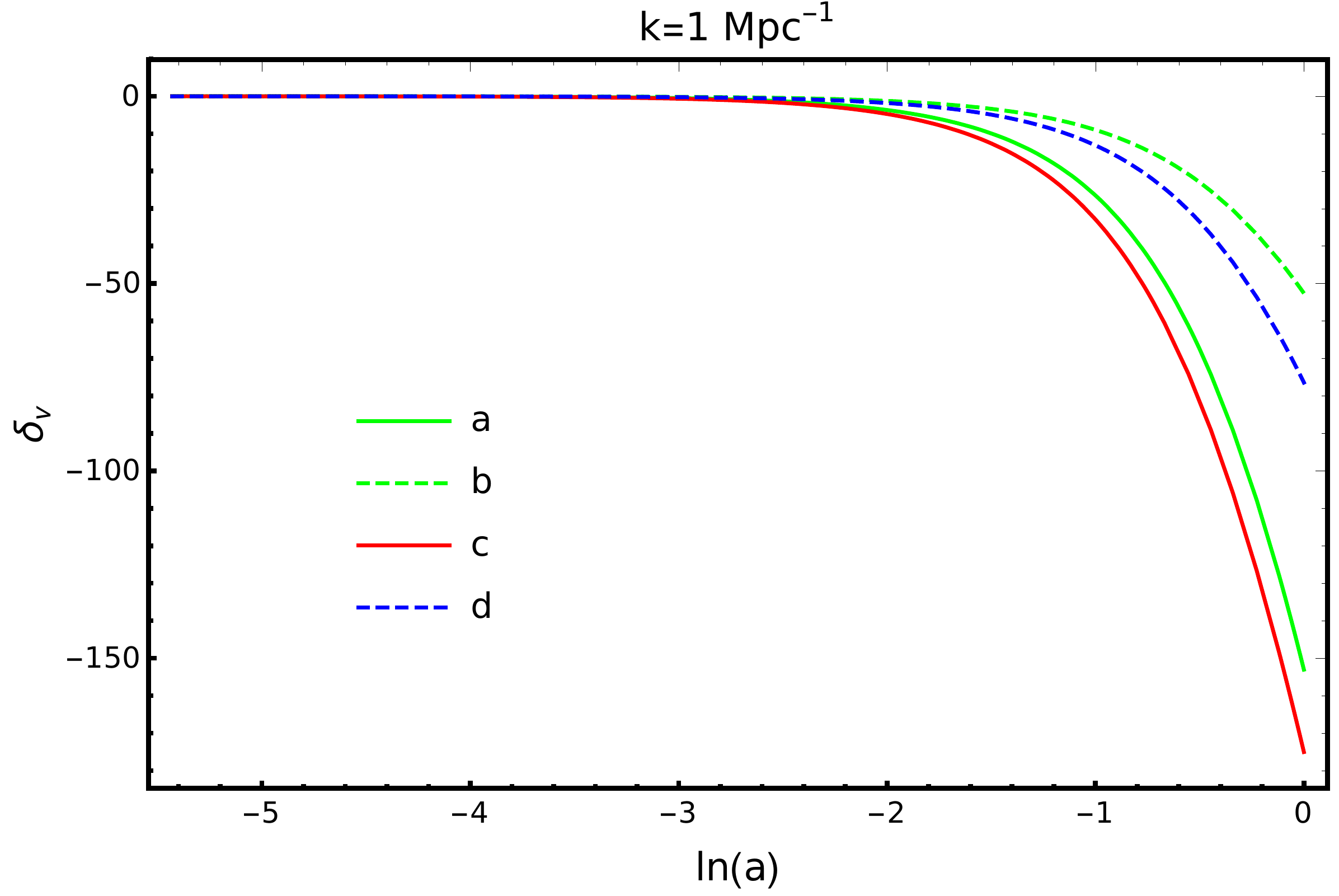} }
    \qquad
    \subfloat{\includegraphics[width= 0.35\textwidth]{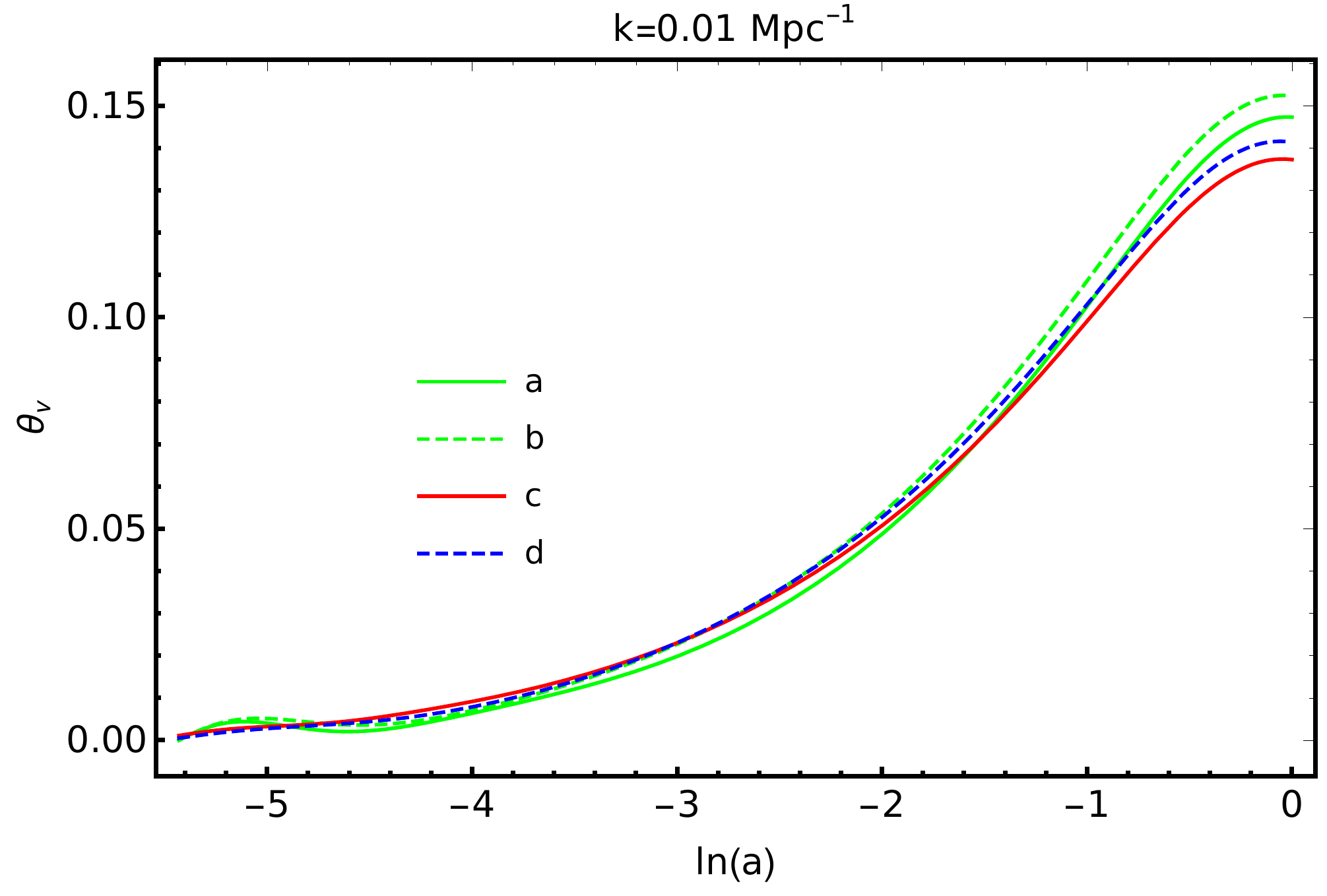} }
    \subfloat{\includegraphics[width=0.35\textwidth]{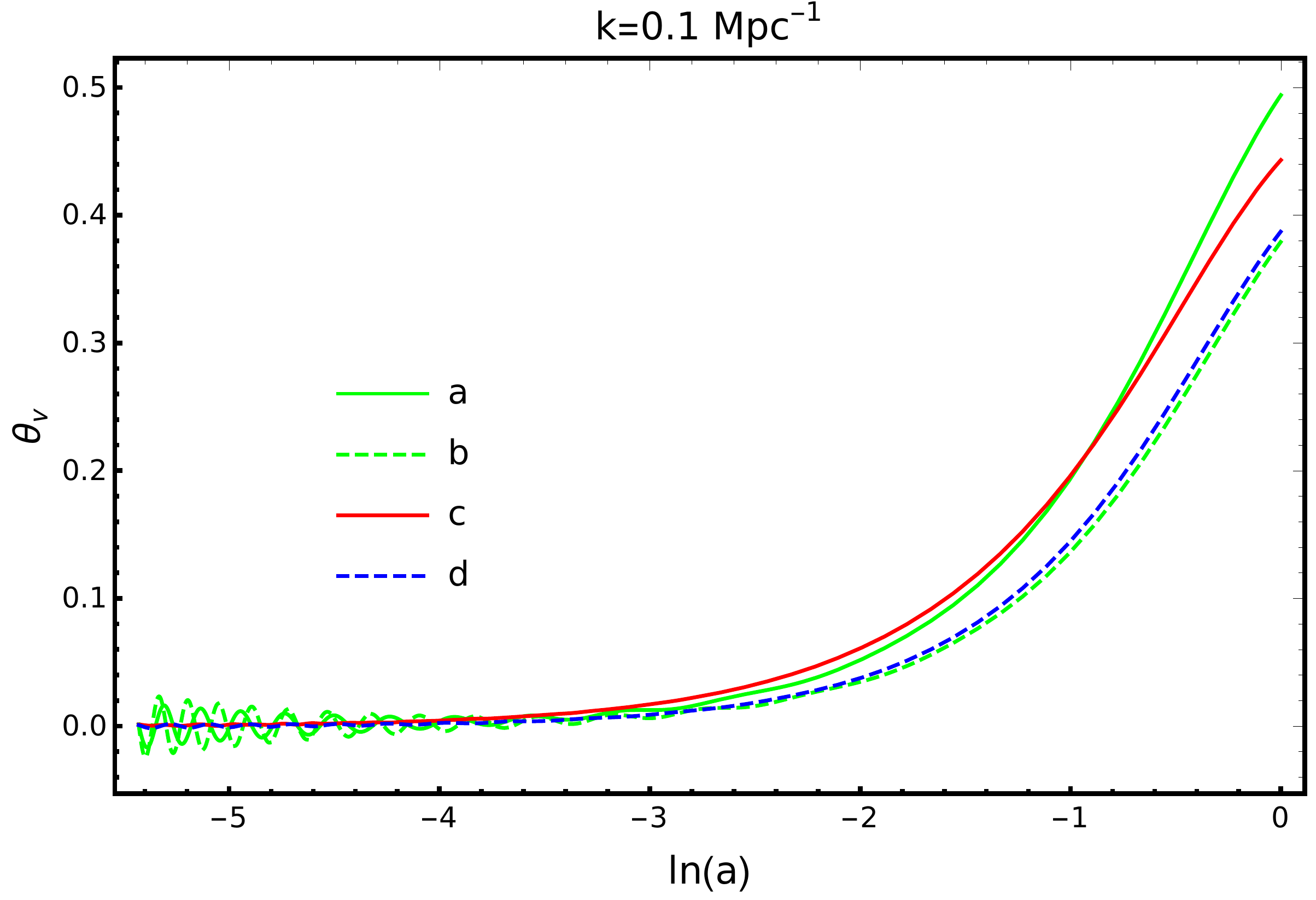} }
    \subfloat{\includegraphics[width=0.35\textwidth]{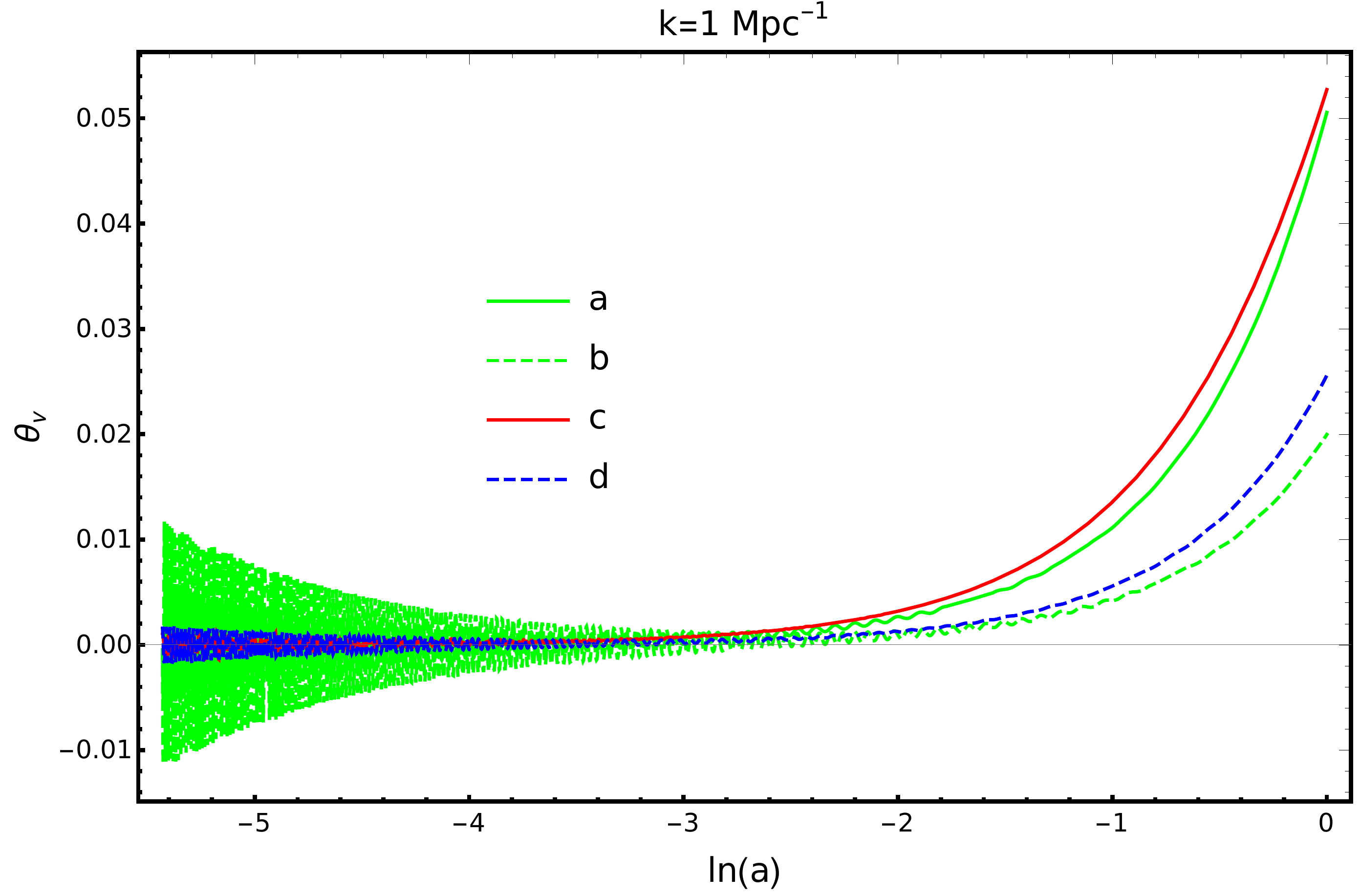} }
    \caption{Neutrino density contrast and velocity divergence as a function of scale factor (starting from non-relativistic transition) for a set of modes  ($k = 0.01,\, 0.1,\, 1 \, \text{Mpc}^{-1}$) with neutrino mass $0.12 \;\text{eV}$. Here curve (a) and (b) in solid and dotted green are plotted with $k_0$ values extracted from \texttt{CLASS} full Boltzmann hierarchy and from eq.~(\ref{kfree0}) respectively, as explained in the main text. Whereas curves (c) and (d) in solid red and dotted blue represents \texttt{CLASS} results using full hierarchy and fluid approximation respectively. }
    \label{neutrino_transfer_1}
\end{figure}

We now proceed to analyse the behaviour of the evolution of neutrino perturbation in $\Lambda$CDM background corroborating the simple assumption for $k_{\mathrm{FS}}$ with physically relevant initial conditions.  
We are primarily concerned with the post-transition epoch when neutrinos behave like matter. In order to verify the validity of the fluid approximation, we first consider the exact solution (Method 1) for neutrino perturbations as described in eq.~(\ref{dnu00order}), supplemented by suitable initial conditions. As mentioned earlier, we choose two different sets of initial conditions for the fluid equations. The first set utilizes the full $k$-dependent neutrino profile at the transition redshift extracted from \texttt{CLASS} code, denoted as ``CLASS IC''. The second set employs an approximation of the same profile through a smoothed Heaviside step function at the transition time, denoted as ``Heaviside IC'', in accordance with the fact that the neutrino perturbations are essentially negligible at and after the transition due to oscillations and traces the CDM at large scales. The ``Heaviside IC'' is constructed as $\delta_{\nu}(k,\tau_{i})= \delta_{\nu}^{\mathrm{CLASS}}(k_{\text{NR}},\tau_{i}) \Theta(k-k_{\mathrm{NR}})$ where $\delta_{\nu}^{\mathrm{CLASS}}(k_{\text{NR}},\tau_{i})$ is extracted from \texttt{CLASS} and $\Theta$ is a smoothed Heaviside function. We aim to investigate the impact of these two different initial conditions (CLASS IC and Heaviside IC) on the neutrino transfer function as a function of wavenumber. In figure\ref{neutrino_transfer_1}, we present the over-density and velocity divergence of non-relativistic neutrinos  for a set of modes ($k = 0.01,\, 0.1,\, 1 \, \text{Mpc}^{-1}$). The figure makes a thorough comparison among four curves (a-d):\\\\(a) The exact solution eq.~(\ref{dnu00order}) of the neutrino perturbation, supplemented by the CLASS IC which includes a $k_0$, obtained with the high-precision settings of \texttt{CLASS} (taking into account the scale dependence of the neutrino sound speed).\\\\ (b) The exact solution eq.~(\ref{dnu00order})  of the neutrino perturbation, supplemented by the CLASS IC and $k_0$ given by eq.~(\ref{kfree0}) i.e. without incorporating scale dependence of neutrino sound speed. \\\\
(c)  Results obtained using \texttt{CLASS} with the full Boltzmann hierarchy extended down to redshift zero.\\\\(d) Results from \texttt{CLASS} using the default switch to the fluid approximation at low redshift. \\\\  
 We have used the cosmological parameters,  $\Omega_{\mathrm{m,0}} = 0.3111,\; \Omega_{\Lambda,0}= 0.6889,\; \Omega_{\mathrm{b,0}} = 0.04$ for all the plots and \texttt{CLASS} settings\footnote { These are the high precision settings we employed in \texttt{CLASS}: $\mathbf{ncdm\_fluid\_approximation} = 3 $ (to switch off the CLASS default fluid approximation), $\mathbf{l\_max\_ncdm} = 30 $ } for neutrinos are mentioned in the footnote. It follows from the analysis of a range of modes in figure \ref{neutrino_transfer_1} that the neutrino density and velocity perturbations, as adapted in our fluid model, i.e. curve (a) closely match the results derived from \texttt{CLASS} when employing the full Boltzmann hierarchy i.e. curve (c). Also let us note that  the curve (b) generated with $k_0$ value as derived from eq.~\eqref{kfree0} closely traces the curve (d) generated by \texttt{CLASS} fluid approximation rather than curve (c) that utilizes the full Boltzmann hierarchy.

\begin{figure}[ht!]
    \centering
    \includegraphics[width=0.65\textwidth]{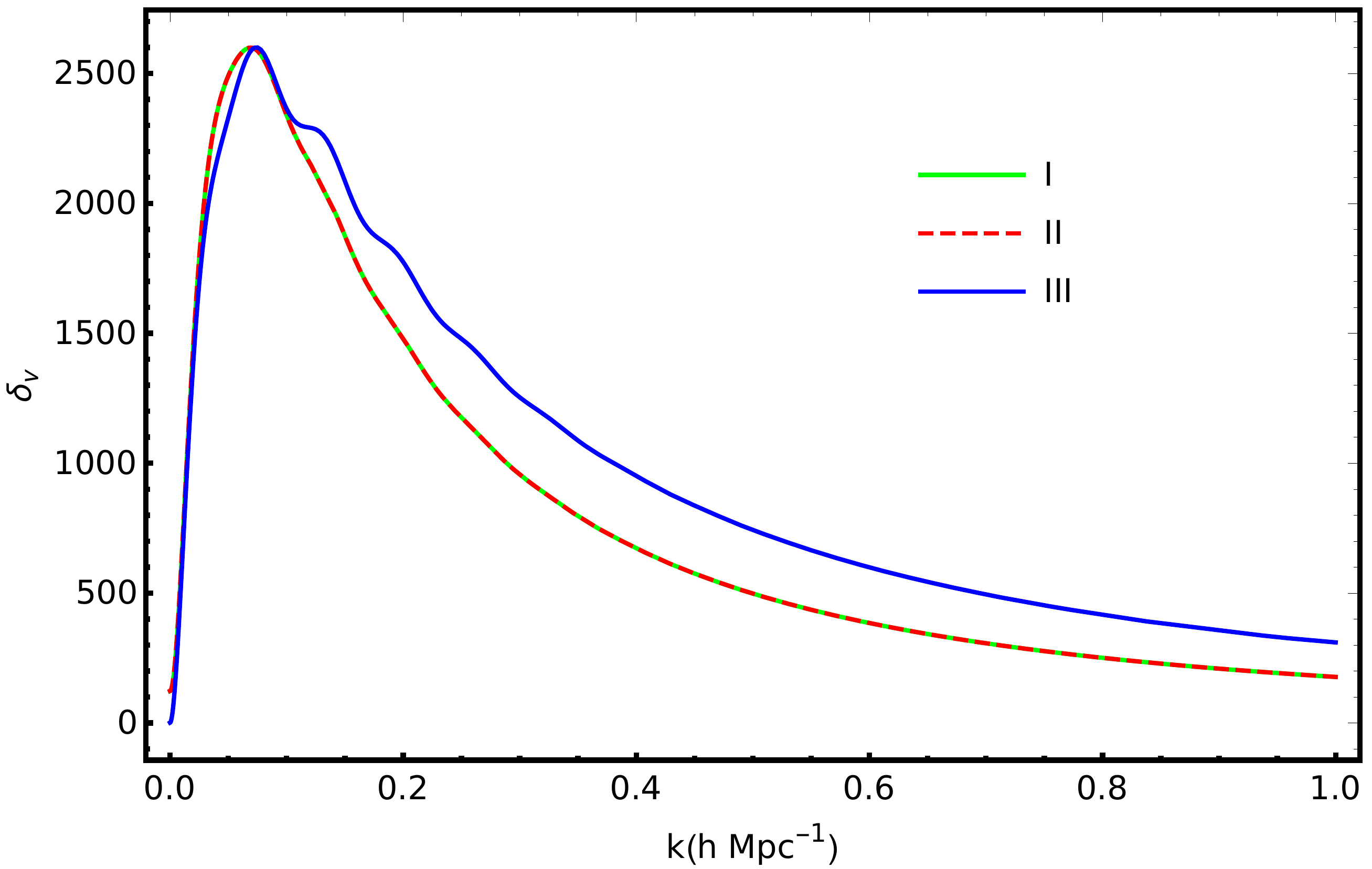} 
    \caption{   Absolute value of neutrino density contrast as a function of $k$ with initial condition extracted from CLASS denoted as CLASS IC (curve I) and with smoothed Heaviside step function denoted as Heaviside IC (curve II). The blue solid curve (curve III ) represents the exact solutions from CLASS in high precision settings in $\Lambda$CDM background.}
    \label{neutrino_transfer_2}
\end{figure}

We have assessed the influence of the two different initial conditions in figure~\ref{neutrino_transfer_2}, i.e.  ``CLASS IC'' and ``Heaviside IC''. Here the 3 different curves arise in figure~\ref{neutrino_transfer_2} from the following considerations:\\\\
(i) represents neutrino density contrast evaluated from the exact solution eq.~(\ref{dnu00order}) (Method 1) using CLASS IC i.e. the full k dependent profile of both CDM and neutrino imported from CLASS at the transition redshift as initial conditions.\\\\ (ii) represents the same profile with CLASS IC for CDM perturbation and Heaviside IC for the neutrino solution eq.~(\ref{dnu00order}).\\\\  (iii) represents CLASS results using full Boltzmann hierarchy down to redshift zero.\\\\
It is important to note that, at small scales, the fluid approximation exhibits notable deviations from the solutions provided by \texttt{CLASS} using the full Boltzmann hierarchy. This difference is anticipated since for this plot we have chosen to ignore the scale dependence of the free-streaming length entering via the neutrino sound speed.

\subsection{Impacts on matter power spectrum}
\label{suppression in the power spectra}

\begin{figure}[!h]
    \centering
    \subfloat{\includegraphics[width= 0.45\textwidth]{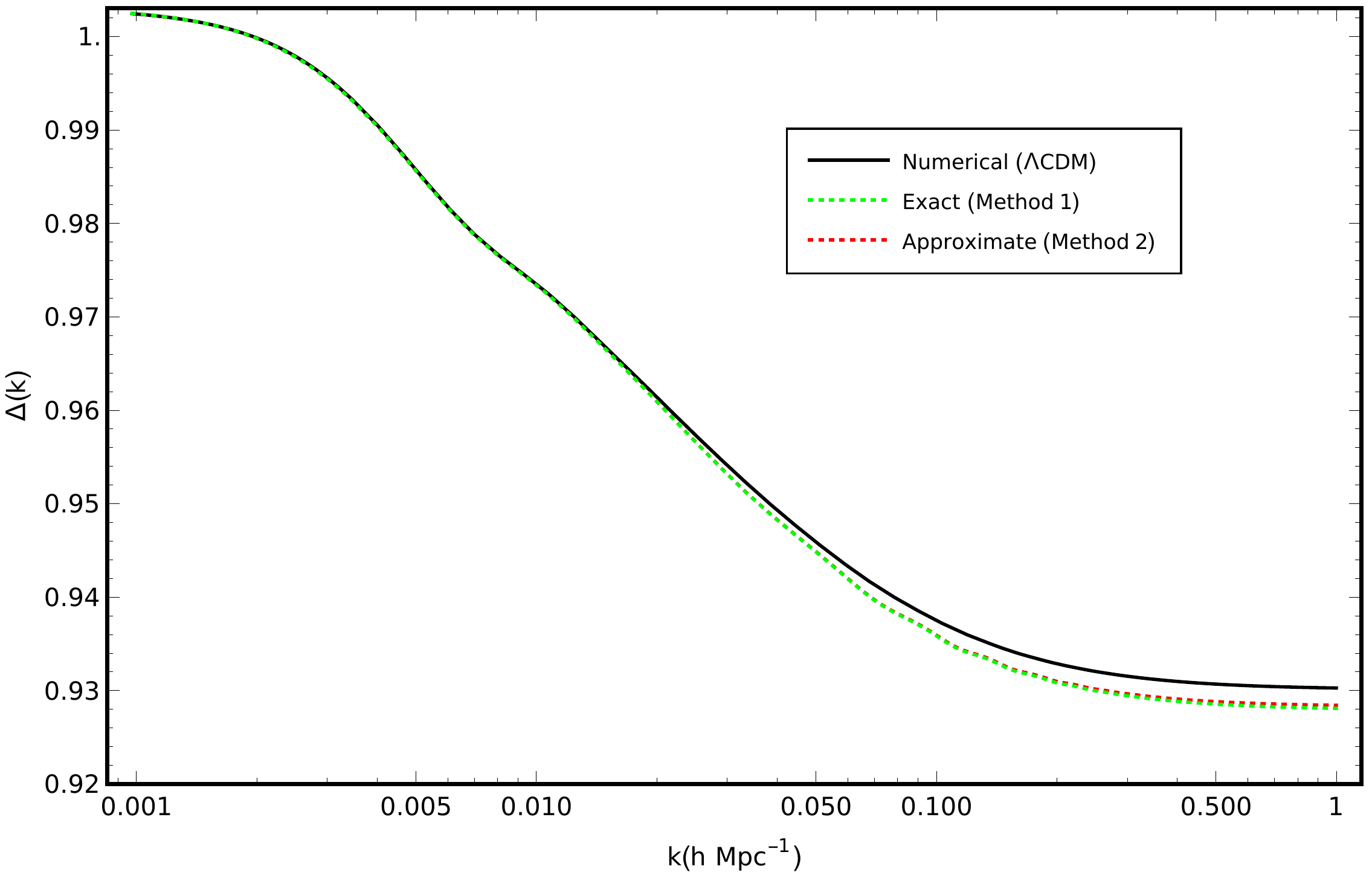} }
    \qquad
    \subfloat{\includegraphics[width= 0.45\textwidth]{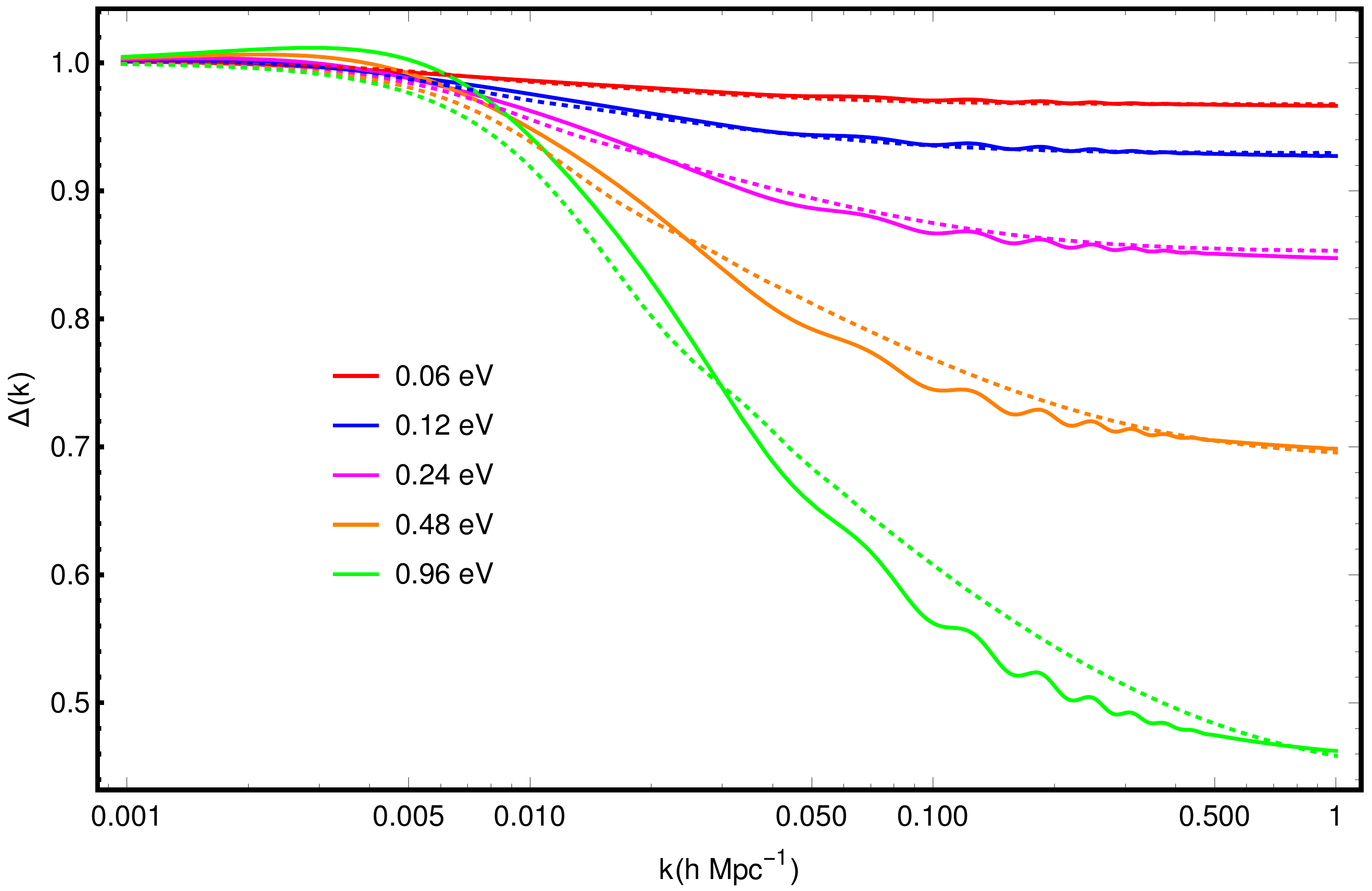} }
    \caption{\textbf{Left Panel:} We have shown the ratio of the matter power spectrum in presence of massive neutrinos with $\sum{m_{\nu}}=0.12$ eV (corresponding $f_\nu = 0.009$) to the power spectrum with $f_\nu=0$. Here the dotted and solid curves are respectively the analytic  solutions (following two different methods) and numerical results of the coupled two fluid equations in $\Lambda$CDM background. \textbf{Right Panel:} Here the solid curves are plotted using \texttt{CLASS} high precision settings and the dotted curves represent the numerical solutions for different $\sum m_{\nu}$ in $\Lambda$CDM background.}
    \label{fig:analytical}
\end{figure}

As usual, the total matter power spectrum in the Fourier space is defined as,
\begin{align}
    \left<\delta_{\rm{m}}(\vec{k}_1,\tau)\delta_{\rm{m}}(\vec{k}_2,\tau) \right> = (2\pi)^3 \delta_{\rm D}^{[3]}(\vec{k}_1-\vec{k_2})P(k,\tau)
\label{PS}
\end{align}
where $\delta_{\rm D}^{[3]}$ is the 3D Dirac delta function. The matter power spectrum on large scales is unaffected since the neutrino perturbations  behave like CDM in this limit. However, on small scales the power spectrum is  modified according to the definition of $\delta_{\rm m}(k,\tau)$ in eq.~\eqref{matterDP}. Therefore the matter power spectrum at any time $\tau$ can be expressed as,
\begin{align}
    P(k,\tau) = &\left(f_\nu \delta_{\nu}^{(0)}(k,\tau)+(1-f_\nu)(\delta_{\rm cb}^{(0)}(k,\tau)+\delta_{\rm cb}^{(1)}(k,\tau))\right)^2.
\label{powerlo}
\end{align}
where $\delta_{\rm{cb}}^{(0)}$ is given by  eq.~\eqref{0orderdcb} and $\delta_{\nu}^{(0)}$ and $\delta_{\rm{cb}}^{(1)}$ are  presented in eqs.~(\ref{dnu00order}) and \eqref{dcbwi1} for Method 1 and  eqs.~(\ref{deltanu0}) and (\ref{deltacb1order}) for Method 2. For the purpose of the present study, we choose the initial condition for neutrino perturbation to be full k-dependent profile of neutrinos at the transition redshift extracted from  \texttt{CLASS}.
For the CDM sector the initial condition considered is the full k-dependent CDM profile at the transition redshift. On very large scale as one should recover $P(k,\tau) = P_{f_{\nu}=0}(k,\tau)$ where $P_{f_{\nu}=0}(k,\tau)$ is the power spectrum in the absence of neutrino perturbation, so any non-trivial effect of neutrino perturbations on matter power spectrum can be identified by a quantity $\Delta(k,\tau)$ defined as follows,
\begin{align}
   \Delta(k,\tau) \equiv \frac{P_{f_{\nu}\neq 0}(k,\tau)}{P_{f_{\nu}=0}(k,\tau)} .
\label{RPS}
\end{align}
In what follows, we will be primarily interested in computing $\Delta(k,\tau)$ at present time $\tau=0$.  
We have examined the validity of the analytical  eqs.~(\ref{dnu00order}) and \eqref{dcbwi1} obtained via Method 1  and also the corresponding ones  eqs.~(\ref{deltanu0}) and (\ref{deltacb1order}) obtained via Method 2.
For further comparison, we have also solved the coupled two fluid equations (\ref{dcb''}) and (\ref{dnu''}) numerically, to all orders in $f_{\nu}$. Subsequently, we compared these numerical results with the analytical results, which are valid up to the first order in $f_{\nu}$ for $\Lambda$CDM and $w$CDM universe. In the left panel of figure~\ref{fig:analytical}, we have compared the numerical solutions to the two fluid system with the exact solution (Method 1) and approximate analytic solution (Method 2) in $\Lambda$CDM background considering $k_0$ as indicated by eq.~(\ref{kfree0}). Also we have considered the full k-dependent profile of both CDM and neutrinos at the transition redshift as the initial condition (CLASS IC). 
The left panel of figure~\ref{fig:analytical} shows that the analytical solutions following Method 1 and Method 2 for the matter power spectrum are in remarkable agreement for a wide range of k values in the linear regime.
In the right panel of figure~\ref{fig:analytical}, we have shown the suppression of matter power spectra for different neutrino masses resulting from the numerical solutions (correct to all orders in the neutrino fraction) and \texttt{CLASS} full Boltzmann hierarchy solutions. The comparison reveals a significant level of agreement between our results  and those from \texttt{CLASS} up to a specific critical value of neutrino mass, well within the current experimental limits. However, as we increased the neutrino mass parameter to higher values, our results and those from \texttt{CLASS} start to exhibit slight to moderate inconsistencies.

To summarize, the exact solution (Method 1) eqs.~(\ref{0orderdcb} - \ref{dcbwi1}) in subsection \ref{solutions} and the approximate solution (Method 2) eqs.~\eqref{deltanu0} and \eqref{deltacb1order} in subsection \ref{method2} allow us to construct the suppression of total matter power spectrum in presence of massive neutrinos in $\Lambda$CDM universe and beyond. Our results may be looked upon as a building block to incorporate the effects of DE at mildly nonlinear regime which is also important for momentum conservation, see \cite{Blas:2014hya,Garny:2022fsh} for detailed study. We also expect that our analytical result will lead to improvements in the transfer function in the mildly nonlinear regime as opposed to invoking halofit model in the simulation. In the nonlinear description of fluid models with massive neutrino, the nonlinear density contrast of neutrino sector is often approximated through
$\delta_{\rm{\nu}}\sim \delta_{\rm{cb}} \times  \frac{\delta_{\nu}^{\rm {lin}}}{\delta_{\rm{cb}}^ {\rm {lin}}}$ as mentioned in \cite{Blas:2014hya,Garny:2020ilv}. We expect that the analytical results of neutrino and CDM perturbations as obtained above will help to better approximate the nonlinear density contrast in this two fluid model scenario. Moreover, a combination of analytic approach together with halofit (or any such simulation) has great potential to yield a more accurate power spectrum at yet unexplored length scales. This is expected to leave some imprints on the future LSS missions and would be an interesting aspect to explore in the era of precision cosmology.   

\subsection{Discussions on velocity spectrum}

The velocity divergence of the CDM (+ baryon ) sector is related to the corresponding density profile by a factor of $\mathcal{H}f$, where $f$ is the growth rate of the CDM sector. However, in the presence of massive neutrinos coupled with CDM perturbations, the growth rate exhibits a non-trivial dependence on wave number resulting in a suppression of the velocity divergence spectrum, even in the linear regime. The velocity divergence spectrum of CDM in Fourier space is defined as,
\begin{align}
    \left<\theta_{\rm m}(\vec{k}_1,\tau)\theta_{\rm m}(\vec{k}_2,\tau) \right> = (2\pi)^3 \delta_{\rm D}^{[3]}(\vec{k}_1-\vec{k_2})P^{\theta\theta}(k,\tau)
\label{VS}
\end{align}
where $\delta_{\rm D}^{[3]}$ is the 3D Dirac delta function.
To explore this above-mentioned feature, we have plotted the velocity spectrum of the CDM sector in the presence of massive neutrinos using a two-fluid scenario and found good agreement with the results obtained from the \texttt{CLASS} code, as long as $f_{\nu} \ll1$ (satisfying the neutrino mass limit).
\begin{figure}[ht!]
 \centering
 \includegraphics[width= 0.45\textwidth]{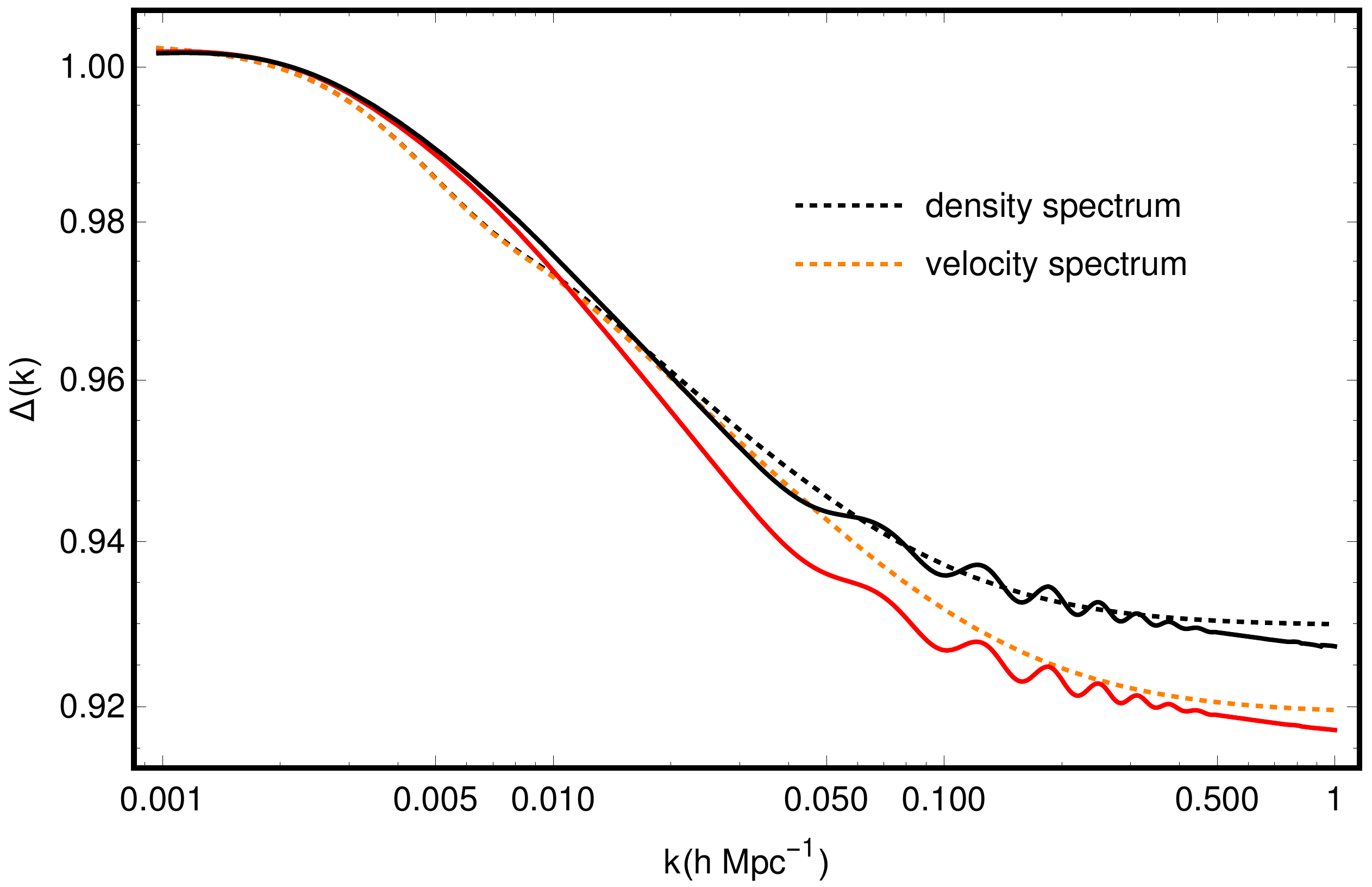}
 \caption{Comparison of analytical solutions for density and velocity power spectrum with \texttt{CLASS} in $\Lambda$CDM background for $\sum{m_{\nu}}=0.12$ eV. Here the orange curves represent the suppression of the velocity spectrum and the black curves represent the suppression of the density power spectrum.}
   \label{fig:numerical}
\end{figure}
The suppression ratio of the velocity divergence spectrum is defined just like the density profile as follows,
\begin{align}
   \Delta(k, \tau) \equiv \frac{P^{\theta\theta}_{f_{\nu}\neq 0}(k, \tau)}{P^{\theta\theta}_{f_{\nu}=0}(k, \tau)} .
\label{VPS}
\end{align}

Figure~\ref{fig:numerical} presents a comparison of the two-fluid velocity spectrum with the \texttt{CLASS} result alongside the suppression of the density power spectrum. The figure again validates both the analytical (\ref{0orderdcb} - 
 \ref{dcbwi1}) and numerical solutions of the two-fluid model. Further, probing velocity spectrum with upcoming surveys may lead to new insights into structure formation with massive neutrinos as reported in some recent N-body simulation results \cite{Zhou:2021sgl}. We leave such detailed investigations for future works.

\section{CDM-massive neutrino perturbations in beyond-\texorpdfstring{$\Lambda$}{}CDM backgrounds}
\label{Two-fluid equations in Different DE Universe}

As we mentioned earlier, effects of massive neutrinos on matter power spectrum in  $\Lambda$CDM background and beyond come through scale- and time-dependent sound speed of neutrinos, which make the system more interesting than vanilla $\Lambda$CDM scenario with CDM perturbations only. A detailed analysis of the two-fluid perturbations in dynamical DE background will  help  us to enrich our understanding of the interplay between CDM and neutrino perturbations at different scales. This will further help in a better understanding of the transfer function with more theoretical inputs and could also help in numerical simulations. This is expected to have yet unexplored far-reaching consequences on nonlinear evolution too. 

Let us briefly review the dynamics of a few DE backgrounds that we will make use of in this section. As is well-known, apart from $\Lambda$CDM, there have been several attempts to describe the DE component via phenomenologically motivated models. This is particularly important today since the baseline $\Lambda$CDM model has been found to suffer from a few tensions at varied extent, like the $H_0$ and $\sigma_8$ tensions. As recently been studied in a series of papers (see, for example, \cite{RoyChoudhury:2018gay, DiValentino:2021izs, Shah:2023rqb}), different DE parametrizations to some extent can potentially shed light on tensions involving cosmological parameters. Thus, any non-trivial effects of neutrino perturbations on CDM in these backgrounds need to be well-studied both analytically and with quantitative estimates in order to have a better understanding of perturbations as well as specific DE models/parametrizations in the background. Even though the effects turn out to be small, this is a very crucial aspect, especially in the era of precision cosmology. 

The (time-varying) equation of state of the DE component is defined as $w(z)\equiv \frac{p_{\rm DE}}{\rho_{\rm DE}}$, which reflects on the Friedmann equation in the following way, 
\begin{align}
    {\mathrm{H^2}}/{\mathrm{H}_{0}^2}= \Omega_{\rm m,0}(1+z)^3 +\Omega_{{\rm DE},0} ~f(z)
\end{align} where the effects of the Equation of State (EoS) of DE is encapsulated in the function,
\begin{align}
    f(z)=e^{ \left[3\int_{0}^{\ln(1+z)}(1+w(z^{\prime}) \, d\ \ln(1+z^{\prime})\right]}.
    \label{f(z)}
\end{align}
Here $\Omega_{\rm m,0}$ and $ \Omega_{\rm DE,0} $ are the present day matter (cold dark matter+baryons+massive neutrinos) and DE density respectively, satisfying $\Omega_{\rm m,0}+\Omega_{\rm DE,0}=1$.
We now proceed to investigate the two fluid model in different DE backgrounds  and illustrate the effects of massive neutrinos on the total matter power spectrum.

 \subsection{DE with constant EoS: \texorpdfstring{$w$}{}CDM background}
 
Perhaps the simplest possible scenario beyond  $\Lambda$CDM is to consider a $w$CDM model where the DE equation of state $w$ is redshift-independent. Here, for the case of any constant $w$, eq.~\eqref{f(z)} simply turns out to be,
 \begin{align}
     f(z)=(1+z)^{3(1+w)}.
 \end{align}
 where $w=-1$ case behaves like a cosmological constant for which the solution is given by eqs.~(\ref{0orderdcb} - \ref{dcbwi1}). For other values of $w$, one can obtain an exact solution similar to eqs.~(\ref{0orderdcb} - \ref{dcbwi1}) i.e. Method 1. However, for brevity, we only present  the approximate solution in $w$CDM universe analytically following Method 2 as mentioned in subsection \ref{method2}, without providing the rather lengthy expressions. These results are presented in figure~\ref{Figure5} where we compare the suppression of matter power spectrum obtained via Method 2 with a numerical solution for $w = -1.1$ with the ``CLASS IC'' and $k_0$ obtained from eq.~\eqref{kfree0} in the left panel of figure~\ref{Figure5}. 
 We further compare these results with that obtained from \texttt{CLASS} in the right panel of figure~\ref{Figure5} alongwith other DE backgrounds as well.
\subsection{DE with evolving EoS in the  background}

There is a plethora of DE models with variable EoS, either motivated from a pure phenomenological point of view or emerging from relatively fundamental perspectives, or even from the mere goal of addressing the Hubble tension in recent times. All of them essentially deals with a functional form of the EoS evolving with redshift $w(z)$, called the EoS parametrization. In what follows, we will hand pick two well-known parametrizations as the background and explore the nature of perturbations in those backgrounds.

\subsubsection{CPL parametrization}

Perhaps the most popular DE model after $\Lambda$CDM is the Chevallier-Polarski-Linder (CPL) parameterization \cite{Linder:2002et, Chevallier:2000qy} also known as $w_{0}w_{a}$CDM parametrization. It is a two-parameter extension to $\Lambda$CDM with a redshift dependent equation of state given by,
\begin{align}
    w(z)=w_0 +w_a\frac{z}{1+z}
\end{align}
Planck 2018 constrains the parameters $w_0$ and $w_a$ in the CPL parameterization to be $w_0={-1.21}^{+0.33}_{-0.60} $ and $w_a < -0.85$ at $68\%$ CL \cite{Planck:2018vyg}, where $w_0$ represents the equation of state today and $w_a$ describes its time evolution. 

\subsubsection{JBP parametrization}

More generically one can proceed to construct a class of CPL-like parametrization where the redshift-dependent part scales as $\frac{z}{(1+z)^p}$, with $p$ being a natural number. 
The Jassal-Bagla-Padmanabhan (JBP) parametrization \cite{Jassal:2005qc} is one such example in this family with $p = 2$, i.e. it proposes a DE equation of state of the following form, 
\begin{align}
    w(z)=w_0 +w_a\frac{z}{(1+z)^2}.
    \end{align}   
In redshift-dependent DE backgrounds like CPL and JBP as described above,
\begin{figure}[ht!]
    \centering
    \subfloat{\includegraphics[width= 0.45\textwidth]{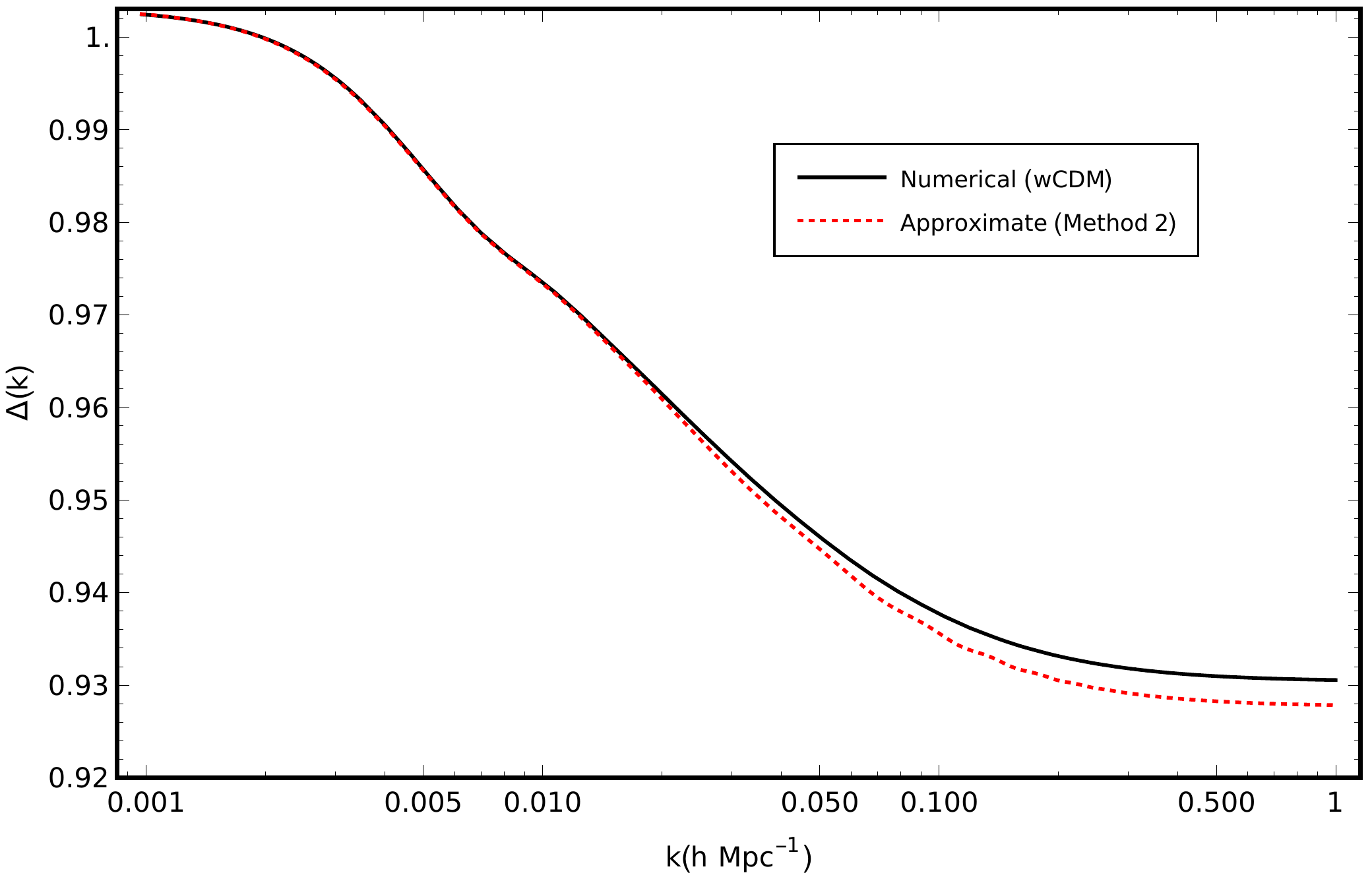} }
    \qquad
    \subfloat{\includegraphics[width=0.45\textwidth]{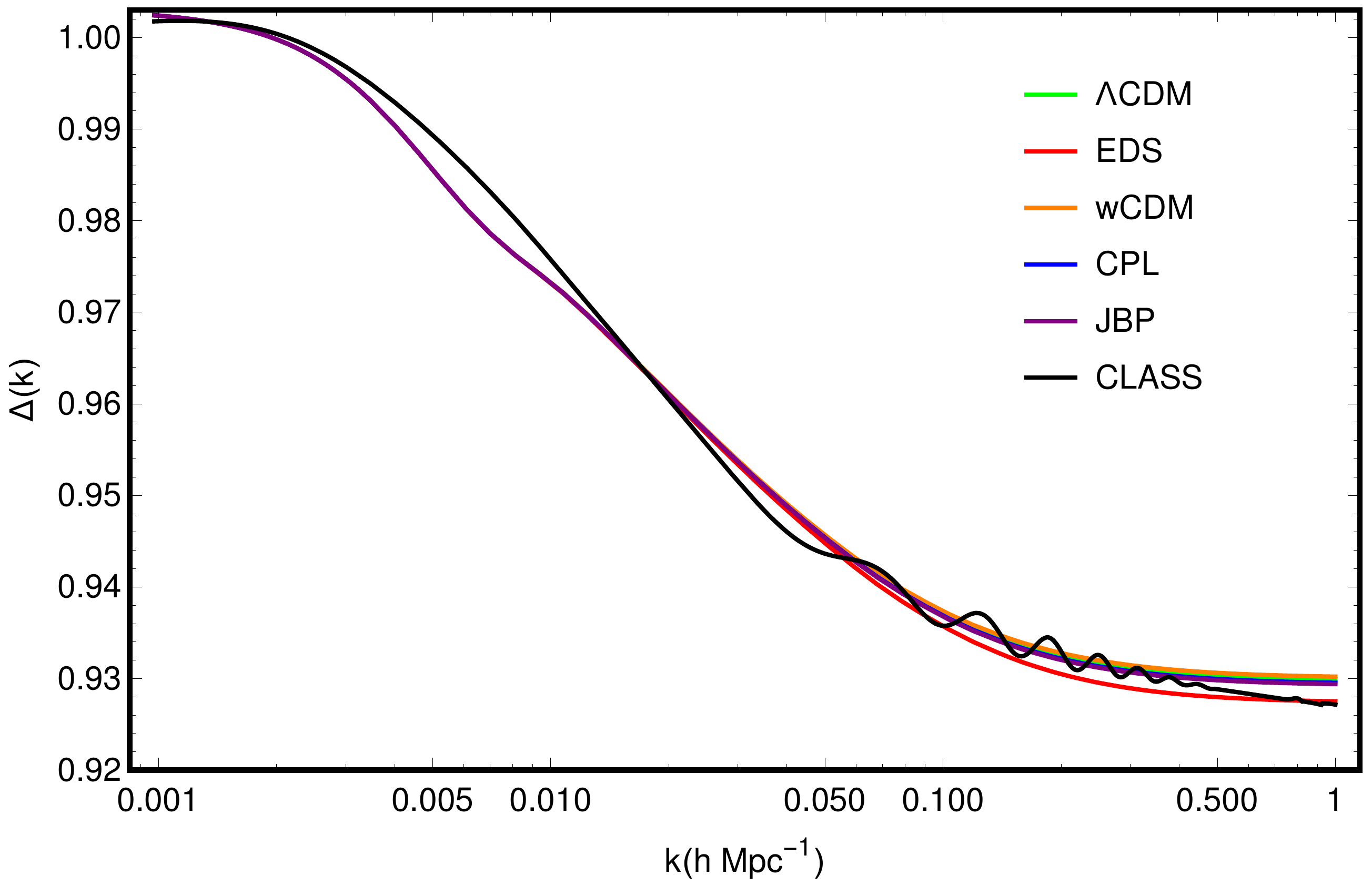} }
    \caption{Suppression of matter power spectrum in different DE backgrounds. The \textbf{Left Panel} indicates a comparison of the matter power spectra between numerical and analytical solution (via Method 2) in $w$CDM cosmology and the \textbf{Right Panel} shows a comparison between \texttt{CLASS} solution and numerical solutions in different DE backgrounds.}
    \label{Figure5}
\end{figure}
the factor $(1+\frac{{\mathcal{H}}^{'}}{\mathcal{H}})$ and fractional matter density $\Omega_{\rm m}(\tau)$ in eqs.~\eqref{dcb''} and \eqref{dnu''} respectively take the form,
\begin{eqnarray}
   1+\frac{{\mathcal{H}}^{'}}{\mathcal{H}}=2-\frac{3}{2}\frac{1+\frac{Q}{3}e^{3\tau}\frac{df}{d\tau}}{1+Q e^{3\tau}f(\tau)}, ~\Omega_m(\tau)= \frac{1}{1+Q e^{3\tau}f(\tau)}
\end{eqnarray}
where $Q$ is defined as $\frac{\Omega_{\rm DE}}{\Omega_{\rm m}}$. It can be readily checked that both of them boil down to constant $w$CDM results for $w_0 = w, ~ w_a =0$ and further to $\Lambda$CDM results for $w_0 =-1, ~ w_a =0$.

We can now solve the coupled two fluid equations for CDM and massive neutrinos considering these different DE backgrounds. The results have been plotted in the form of suppression of power spectrum in figure \ref{Figure5} and also compared  with that obtained from \texttt{CLASS}. 

Figure\ref{Figure5} summarizes the results for all the cosmological models under consideration in the present article alongside EDS. As evident from the left panel of figure\ref{Figure5}, there is $\sim 0.2\%$ difference in suppression between analytical and numerical results in $w$CDM universe,  whereas the difference between dynamical DE models is negligible in linear regime even with the most optimistic deviation from $\Lambda$CDM (all computations are performed for sum of neutrino mass 0.12 eV which corresponds to $f_\nu =0.009$). Thus, different DE models are expected to remain indistinguishable even after the inclusion of the effects of neutrino perturbations on CDM, so far as the linear scales are concerned. Nevertheless, the fact that the inclusion of neutrino perturbations indeed affects the CDM perturbations, as being reflected on $\Lambda$CDM case is an interesting outcome of the plots and hence of the present analysis. In addition, as argued earlier, the results from the linear scales are going to affect the (mildly) non-linear scales and transfer function, which has the potential to provide useful information for future missions. We hope to address some of these issues in a follow-up paper.

\section{Prospects in observations}
\label{Making Contact With Observations}

Both early and late-time cosmological observations are sensitive to the sum of neutrino masses along with the six standard parameters in the vanilla $\Lambda$CDM model. Since in CMB measurements, the optical depth $\tau$ is degenerate with the amplitude of temperature anisotropy, finding out independent and stringent constraints on the sum of neutrino mass from CMB missions alone is rather challenging \cite{CMB-S4:2016ple}. Improved measurements of $\tau$ from future CMB observations could lead to better constraints on $\sum{m_{\nu}}$. Otherwise, the minimum uncertainty in $\sum{m_{\nu}}$ would be limited to $\sigma(\sum{m_{\nu}})> 0.02 $ eV. Proposed CMB experiments like LiteBIRD \cite{Smith:2016lnt} and PICO \cite{Alvarez:2019rhd}, targeted to ameliorate $\tau$ measurements, could reduce the uncertainty in $\sum{m_{\nu}}$ to $0.009-0.015$ eV \cite{Abazajian:2019eic}. 

In addition to CMB, optical depth can be separately constrained from 21-cm observations. At present the only global 21-cm signal from EDGES data \cite{bowman2018absorption}  does not put any impressive constraint on the value of $\tau$. Reionization physics and simulations can put some kind of theoretical bounds on $\tau$, which can give rise to a slightly different value based on the model of reionization under consideration \cite{Mitra:2019rzc,Hazra:2019wdn}. However, these theoretical bounds need to wait till the real data arrives. Cross-correlation of CMB data with future 21-cm mission \href{https://www.skao.int/en}{SKA} \cite{Weltman:2018zrl} may improve the bounds on $\tau$, which may in turn help us reduce the uncertainties on $\sum{m_{\nu}}$ as well.

On the other hand, the amplitude of the matter power spectrum, which is determined through observations of CMB or galaxy lensing, is directly linked to the total matter density parameter $\Omega_m h^2$. The reduction in power due to the presence of massive neutrinos can be attributed to the mismatch between the lensing amplitude (which characterizes small-scale fluctuations) and the homogeneous matter distribution manifested by the universe's expansion rate. Presently, the evaluation of neutrino mass constraints via BOSS BAO \cite{BOSS:2016hvq} measurements is restricted by the uncertainty in the $\Omega_m h^2$ determination. However, DESI BAO \cite{DESI:2016fyo} is expected to significantly improve this measurement so that the constraint on $\sum{m_{\nu}}$ would presumably be mostly degenerate with the constraints on $\tau$. Clubbing that with future CMB and/or 21-cm missions may help in improving the constraints on $\sum{m_{\nu}}$ by constraining $\tau$.

\begin{figure}[ht!]
 \centering
 \includegraphics[width= 0.45\textwidth]{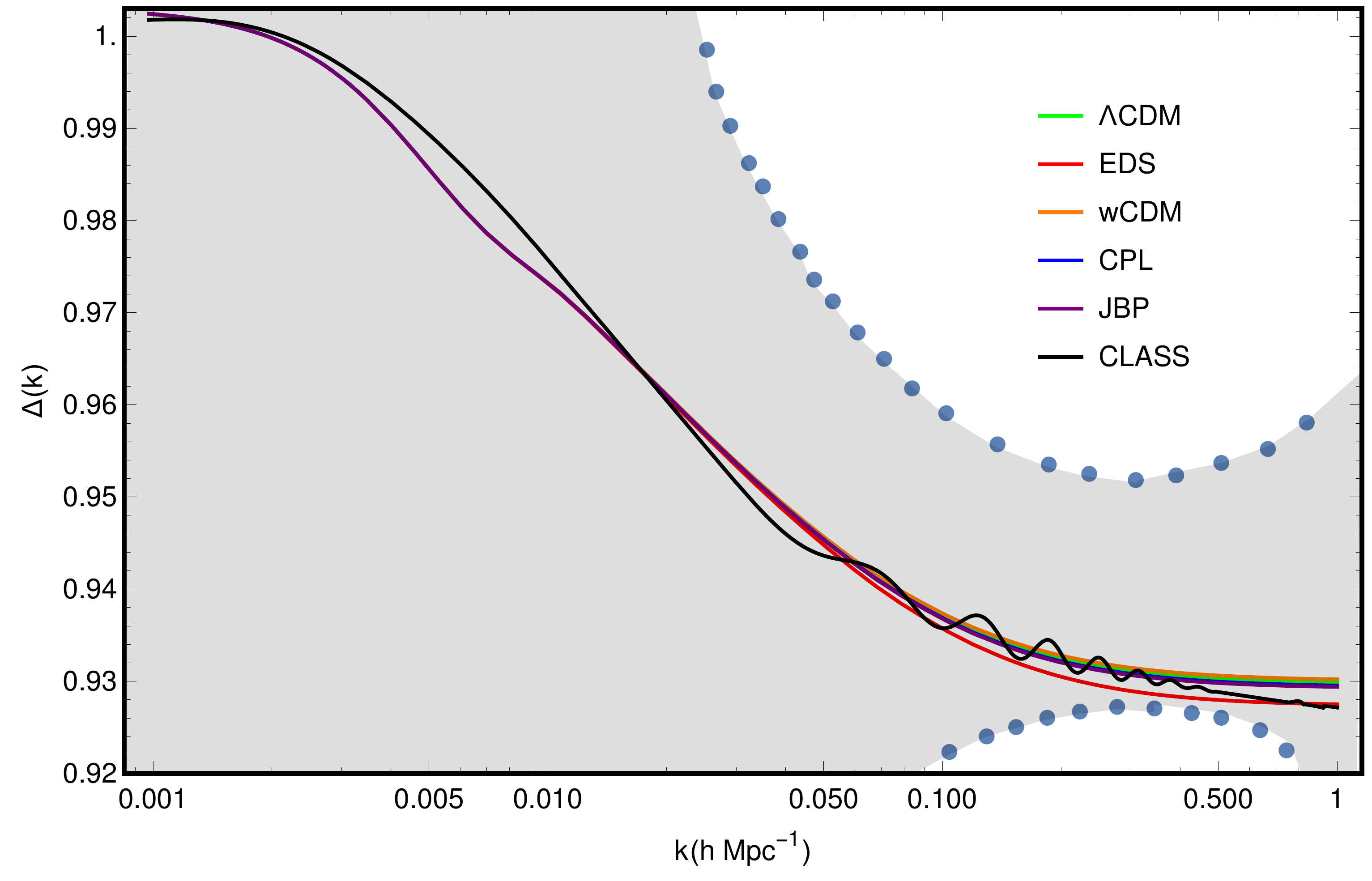}
 \caption{Comparison of 1$\sigma$ noise of BOSS-like galaxy survey and suppression of matter power spectrum in different DE backgrounds  obtained for the present analysis.}
   \label{Figure6}
\end{figure}

With this in mind, we have done a preliminary test for the validity of our analysis  in figure \ref{Figure6}. In this figure, the solid lines represent the theoretical estimate of the suppression of power spectrum in presence of massive neutrinos in different cosmological backgrounds as obtained from figure \ref{Figure5}. We have found that the results closely follow the $\Lambda$CDM scenario even with the most optimistic deviation allowed from Planck \cite{Planck:2018vyg} and BAO observation \cite{BOSS:2016wmc}. Hence, the results would naturally be  quite consistent with current datasets. 
In order to validate our claim further, we plot the $1\sigma $ noise of BOSS-like (BOSS galaxy BAO, BOSS Ly$\alpha$ forest, eBOSS galaxy Broadband data etc.) galaxy survey combining cosmic variance and shot noise \cite{Font-Ribera:2013rwa, Green:2021xzn} in figure \ref{Figure6}, represented by the grey patch with blue dotted boundaries.  It can be found from the figure  that the suppression of power spectrum in presence of massive neutrinos in different  backgrounds lies within the $1\sigma$ uncertainty of BOSS-like galaxy survey noise for $\sum{m_{\nu}} = 0.12$ eV (correspondingly, $f_\nu = 0.009$). This leads to the conclusion that the two-fluid model of CDM and massive neutrino in different cosmological backgrounds presented here are in good agreement with current observations, and would also fall within the $1\sigma$ bounds of a broad class of future LSS surveys.

This calls for a detailed forecast analysis with different future LSS surveys, their combinations, and possible joint analysis with existing CMB and future CMB and/or 21-cm missions. 
As already mentioned, $\sum{m_{\nu}}$ is degenerate with the optical depth $\tau$ in the CMB measurement, and in weak lensing galaxy survey it is degenerate with $\Omega_{\rm m}$.
Future LSS missions like EUCLID \cite{Euclid:2011,Chudaykin:2019ock}, LSST \cite{LSST:2017ags} individually or jointly with present CMB data from Planck18  or future CMB data like LiteBIRD \cite{Smith:2016lnt}, PICO \cite{Alvarez:2019rhd} etc, or 21-cm mission \href{https://www.skao.int/en}{SKA} are expected to put tighter constraints on the sum of neutrino mass for any (beyond)-$\Lambda$CDM backgrounds that will shed more light on the suppression factor in matter power spectrum caused by neutrino perturbations. We hope that the present analysis, although performed mostly as a theoretical quest for systematic development of the coupled CDM-neutrino framework of perturbations, would be more and more relevant with the improvement of experimental precision, hopefully in the next generation surveys in the context of some of the future CMB, LSS and 21-cm missions mentioned above. 

\section{Summary and outlook}
\label{Discussions and Outlooks}

In this article, we have considered coupled two-fluid perturbation equations of CDM (where CDM essentially denotes CDM + baryon sector) and massive neutrinos in different cosmological backgrounds with redshift-dependent neutrino free-streaming length, and investigated possible effects on CDM perturbations in each case. For the vanilla $\Lambda$CDM model, we have obtained solutions to the coupled two fluid perturbation equations following two separate approaches and subsequently found out the total matter power spectrum to first order in $f_{\nu}$. While the solution arising via Method 1 is exact to leading order in the neutrino fraction, it involves numerical integrals, the solution arising from Method 2 is an approximate but fully analytic  solution. The neutrino-mediated suppression of matter power spectrum obtained from the analytical solution has been found to be in good agreement with numerical solutions  to all orders in $f_\nu$. Further, both the analytical and numerical results for the two-fluid framework in $\Lambda$CDM background have been found to be in good agreement with the results obtained from Boltzmann solver code {\texttt {CLASS}} \cite{2011JCAP...09..032L} by over $95\%$ accuracy depending on the neutrino mass fraction. We have then considered several DE backgrounds beyond-$\Lambda$CDM, in particular the $w$CDM and some well-known DE parametrizations like CPL and JBP; and investigated the effects of massive neutrinos on CDM perturbations for each model under consideration. We have also found approximate analytical results in $w$CDM background following the second approach i.e. Method 2 mentioned in subsection \ref{method2} and presented the results in figure~\ref{Figure5}. We have noticed that within the linear regime, the suppression in matter power spectra obtained using analytical results in both $\Lambda$CDM and $w$CDM universes are consistent with numerical results and the outputs from \texttt{CLASS}.  
We further noticed that for the sum of neutrino mass 0.12 eV, the suppression saturates to the well known result  $\sim 8 f_{\nu}$ at a scale $k_{\rm NL}\sim 0.1\rm h \:\rm Mpc^{-1}$ \cite{lesgourgues_mangano_miele_pastor_2013}. The neutrino-mediated suppression of total matter power spectrum in $w$CDM, CPL and JBP-like DE backgrounds differs from $\Lambda$CDM universe by sub-percent level. We further  validate our analysis by searching for the prospects in observations and found that the inclusion of massive neutrinos in fluid models in various background cosmology is within the $1 \sigma$ uncertainty of the BOSS-like galaxy survey noise for $\sum{m_{\nu}}=0.12$ eV set by current observations. 
  
The present analysis can be extended in several directions. First, we approximated the sound speed of the neutrino fluid in terms of the velocity dispersion of the non-relativistic neutrinos \cite{Shoji:2010hm}. It will be interesting to improve this approximation, as has been proposed in a recent study \cite{Nascimento:2023psl}, and study the effects on the matter power spectrum. We also plan to extend the current study to the mildly nonlinear regime in a perturbative scheme, utilizing the linear kernels derived in this work. As argued in the present article, the results are expected to improve upon the current results due to  more accurate analytical inputs from the linear regime. The two-fluid framework can be further generalized by including DE perturbations and constructing a coupled three-fluid system. This might be a bit tricky and would presumably lead to degeneracies, however it would be interesting to study such scenarios at least from a theoretical perspective as of now. From a phenomenological perspective, it will be interesting to consider neutrino self-interaction \cite{RoyChoudhury:2020dmd, Venzor:2022hql, Taule:2022jrz} as well as dark matter-neutrino interactions \cite{Paul:2021ewd, Mosbech:2020ahp} in the fluid approximation framework. Further, all the results obtained from the theoretical analysis and the corresponding nonlinear corrections can be subjected to experimental tests in the upcoming missions, by doing a forecast analysis on   CMB (eg, LiteBIRD \cite{Smith:2016lnt}, PICO \cite{Alvarez:2019rhd} etc), LSS (like EUCLID \cite{Euclid:2011}, LSST \cite{LSST:2017ags} etc.) and 21-cm missions (like \href{https://www.skao.int/en}{SKA} \cite{Weltman:2018zrl}), either by exploring individual missions or by investigating combined constraints. This may in turn impose further constraints on some of the properties of neutrinos, like the total neutrino mass, DM-neutrino interaction strength, effects on CDM power spectrum at nonlinear scales, and possible impact on other cosmological parameters. However, a detailed investigation in this direction is required before we make any concrete comments on this. We plan to report on some of these analyses in future.\\\\
\textbf{Data Availability:} Most of the numerical results presented in this paper have been obtained using \texttt{Mathematica} and publicly available code {\texttt{CLASS}}. The Mathematica notebook files can be shared with the individuals on reasonable requests.

\section*{Acknowledgements}
We gratefully acknowledge the use of the publicly available code \href{https://github.com/lesgourg/class_public}{\texttt{CLASS}}. SP thanks CSIR for financial support through Senior Research Fellowship (File no. 09/093(0195)/2020-EMR-I). RS acknowledges support from DST Inspire Faculty fellowship Grant no. IFA19-PH231 at ISI Kolkata and the OPERA Research grant from BITS Pilani Hyderabad. SP2 thanks the Department of Science and Technology, Govt. of India for partial support through Grant No. NMICPS/006/MD/2020-21. We thank the anonymous Referee for useful suggestions that helped us to improve our paper significantly.

\bibliographystyle{JHEP.bst}
\bibliography{biblio.bib}

\end{document}